\documentclass{article}[12pt] 
\usepackage{ amsmath, amsthm, amssymb } 
\usepackage{color} 
\usepackage{cite}

\newcommand{\be} { \begin{equation} } 
\newcommand{\ee} { \end{equation} } 

\newcommand{\labbel}[1] { \label{#1} } 
\newcommand{\intq} { \int {\mathfrak D}^d q_1 \,{\mathfrak D}^d q_2 } 
\newcommand{\intQ} { \int {\mathfrak D}^d q } 
\newcommand{\sqrtR} {\sqrt{R_4(b;\,b_1,b_2,b_3,b_4)}} 

\newcommand{\Sunrise}[1]{ 
\mbox{\parbox{2.5cm}{\hspace{0.25cm} 
\begin{picture}(2,1) 
\thicklines 
\put(0.3,0.5){\vector(1,0){0.1}} 
\put(0.5,0.5){\line(1,0){1}} 
\put(0,0.5){\line(1,0){0.5}} 
\put(1.5,0.5){\line(1,0){0.5}} 
\put(1,0.5){\circle{2}} 
\put(0.85,1.12){$m_1$} 
\put(0.85,0.60){$m_2$} 
\put(0.85,0.08){$m_3$} 
\put(0.25,0.7){\makebox(0,0)[b]{$#1$}} 
\end{picture} 
}} 
\hfill} 
\begin{document} 
\setlength{\unitlength}{1.3cm} 
\begin{titlepage}
\vspace*{-1cm}
\begin{flushright}
ZU-TH 26/13\\
November 2013 
\end{flushright}                                
\vskip 3.5cm
\begin{center}
\boldmath
{\Large\bf Schouten identities for Feynman graph amplitudes; \\ 
the Master Integrals for the two-loop massive sunrise graph \\[3mm] }
\unboldmath
\vskip 1.cm
{\large  Ettore Remiddi}$^{a,}$
\footnote{{\tt e-mail: ettore.remiddi@bo.infn.it}} 
and {\large Lorenzo Tancredi}$^{b,}$
\footnote{{\tt e-mail: tancredi@physik.uzh.ch}} 
\vskip .7cm
{\it $^a$  Dipartimento di Fisica,
    Universit\`{a} di Bologna and INFN, Sezione di 
    Bologna,  I-40126 Bologna, Italy}
\vskip .4cm
{\it $^b$ Institut f\"{u}r Theoretische Physik, Universit\"{a}t Z\"{u}rich, 
Wintherturerstrasse 190, CH-8057 Z\"{u}rich, Switzerland} 
\end{center}
\vskip 2.6cm

\begin{abstract}
A new class of identities for Feynman graph amplitudes, dubbed 
Schouten identities, valid at fixed integer value of the dimension $d$ is 
proposed. The identities are then used in the case of the two-loop sunrise 
graph with arbitrary masses for recovering the second-order differential 
equation for the scalar amplitude in $d=2$ dimensions, as well as a 
chained set of equations for all the coefficients of the expansions in 
$(d-2)$. The shift from $d\approx2$ to $d\approx4$ dimensions is then 
discussed. 
\end{abstract}
\vfill
\end{titlepage}                                                                
\newpage

\section {Introduction } 
\begin{figure}[h]
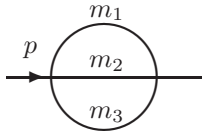

\label{fig:sunrise}
$$ \Sunrise{p} $$ 
\caption{The two-loop sunrise.}
\end{figure}

\labbel{Intro} \setcounter{equation}{0} 
\numberwithin{equation}{section}
The Feynman integrals associated to the two-loop loop self-mass Feynman 
graph of Fig.(\ref{fig:sunrise}),
usually referred to as {\it sunrise}, have been 
widely studied in the literature within the framework of the integration 
by parts identities~\cite{ibpa,ibpb}, and it is by now well known 
that they can be expressed in terms of four Master Integrals (M.I.s), 
\cite{Tarasov1997}, which satisfy a system of four first-order coupled 
differential equations, \cite{Laporta1998} (equivalent to a single 
fourth-order differential equation for any of the Master Integrals). 
Several numerical approaches to the numerical solution of the equations 
with satisfactory degree of precision have been worked out, 
(see for instance \cite{Czyz2009}), but a complete treatment of the general 
case with three different masses in $ d=4 $ dimensions is still missing. \\ 

In the equal mass case the number of 
independent Master Integrals reduces to two, so that the 
two by two first-order system of differential equations can be 
rewritten as a single second-order differential equation for one of the 
Master Integrals, say the full scalar amplitude (see below).  
In Ref.\cite{Laporta2005} it is shown how to build the analytic solution 
of that equation in terms of elliptic integrals, both for $d=2$ and 
$ d=4 $; the two cases, related by the Tarasov's shifting relations 
\cite{Tarasov1996}, are very similar, with the $ d=2 $ case just marginally 
simpler than the $ d=4 $ case. The analytic solution provides with the 
necessary information for writing out very precise and fast converging 
expansions for the accurate numerical evaluation of the two M.I.s
\cite{Pozzorini2006}. 
\par 
More recently, an interesting paper \cite{Weinzierl2012} has shown, by 
using algebraic geometry arguments, that in $d=2$ dimensions 
the full scalar amplitude satisfies a second-order differential equation 
also in the different mass case. The equation was then solved in 
\cite{Weinzierl2013} by suitably extending the method of 
\cite{Laporta2005}; let us observe here that the analytic solution of the 
second-order differential equation is equivalent to the analytic knowledge 
of two (of the four) Master Integrals of the sunrise with different 
masses. \par 
The problem of extending the approach to $ d=4 $, which is the physically 
relevant case, remains, as the straightforward use of the Tarasov's 
dimension-shifting relations is unfortunately not sufficient. 
Indeed, as will be shown in this paper, by explicitly working out the 
shifting relations one finds that any of the four Master Integrals at 
$ d\approx4 $ dimensions can be expressed as a combination of {\it all} the 
four Master Integrals at $ d\approx2 $ dimensions and of the first terms of 
their expansion in $(d-2)$, while the results of \cite{Weinzierl2012} give 
only two of the four Master Integrals at exactly $ d=2 $ dimensions, but 
no other information on the remaining Master Integrals and their 
expansion in $ (d-2) $. \par 
In this paper we introduce a family of particular polynomials in the 
scalar products of the vectors occurring in the Feynman integrals, 
dubbed {\it Schouten polynomials}, which have the property of vanishing at 
some fixed integer value of the dimension $ d $. By using those polynomials 
one can introduce an {\it ad hoc} set of amplitudes, from which one can 
at least in principle extract an independent set of new amplitudes 
which vanish in a non trivial way (see below) at that value of $ d $ 
(say at $d=N$ for definiteness). 
If those new amplitudes are expressed in terms of the previously chosen 
set of Master Integrals, their vanishing gives a set of relations between the 
Master Integrals, valid at $ d=N $, which we call 
{\it Schouten identities}. 
Alternatively, one can introduce a new set of Master Integrals 
including as new Master Integrals some of the independent amplitudes 
vanishing at $ d=N$, write 
the system of differential equations satisfied by the new set of Master 
Integrals and expand them recursively in powers of $(d-N)$ 
around $d=N$. As some of the new Master Integrals vanish at $ d=N $, the 
system of equations takes a simpler block structure. \par 
The pattern is very general, and applies in principle to the integrals 
of any Feynman graph. We work out explicitly the case of the sunrise 
amplitudes at $ d=2 $ with different masses, finding the existence of two 
independent Schouten identities, {\it i.e.} of two independent relations 
between the 
usual Master Integrals at $d=2$, or, which is the same, we can introduce 
a new set of Master Integrals, consisting of two ``conventional'' Master 
Integrals (say the full scalar amplitude and another M.I.) and two new 
Master Integrals vanishing at $ d=2$. 
The system of differential equations satisfied by the new set of Master 
Integrals can then be expanded in powers of $(d-2)$. At zeroth-order 
we find a two by two system for the two ``conventional'' M.I.s 
(the other two Master Integrals vanish), 
equivalent to the second-order equation found in \cite{Weinzierl2012}, 
while at first-order in $(d-2)$ we find in particular two relatively simple 
equations for the first terms of the expansion of the two new M.I.s, 
in which the zeroth-orders of the two ``conventional'' M.I.s appear 
as non homogeneous known terms. \par 
One can move from $d\approx2$ to the physically more interesting 
$d\approx4$ case by means of the Tarasov's shifting 
relations; it is found that for obtaining the zeroth-order term 
in $(d-4)$ of all the four M.I.s (of the old or of the new set) at 
$ d\approx4$ one needs, besides the zeroth-order term in $(d-2)$ 
of the two ``old'' M.I.s at $d\approx2,$ also 
the first term in $(d-2)$ of the new M.I.s. 
\par 
The plan of the paper is as follows: in sec.~\ref{SchPol} we introduce
the Schouten polynomials for an arbitrary number of dimensions, 
while their applications to Feynman Amplitudes is 
discussed in sec.~\ref{SchIds}. In sec.~\ref{NSMI} we show 
how, by using the Schouten Identities,
a new set of Master Integrals can be found, whose differential equations
in $d=2$ take an easier block form and can be therefore re-casted
(see sec.~\ref{IIordDE}) as a second-order differential equation for one 
of the Masters.
In sec.~\ref{Shift} we show how the results at $d\approx4$ can
be recovered from those at $d\approx2$ through Tarasov's shifting relations.
Finally, in sec.~\ref{imMI}, which is somewhat pedagogical, we present 
a thorough treatment of the imaginary parts of the Master Integrals 
in $d=2$ and $d=4$ dimensions.
Many lengthy formulas and some explicit derivations can be found
in the Appendices at the end of the paper.
\section {The Schouten Polynomials} 
\labbel{SchPol} 

As an introduction, let us recall that in $d=4$ dimensions one cannot have 
more than 4 linearly independent vectors; indeed, given five vectors 
$ v_\alpha, a_\mu,b_\nu,c_\rho,d_\sigma $ in four dimensions they are 
found to satisfy the following relation 
\be v_\mu \epsilon(a,b,c,d) - a_\mu \epsilon(v,b,c,d) 
  - b_\mu \epsilon(a,v,c,d) - c_\mu \epsilon(a,b,v,d) 
  - d_\mu \epsilon(a,b,c,v)  = 0 \ , \labbel{MVSchid} \ee 
where $ \epsilon_{\mu\nu\rho\sigma} $ is the usual Levi-Civita tensor 
with four indices, with $ \epsilon_{1234} = 1, etc.$, 
and  following the convention introduced in the program 
SCHOONSCHIP \cite{SCHOONSCHIP} we use 
\be \epsilon(a,b,c,d) = 
 \epsilon_{\mu\nu\rho\sigma}a_\mu b_\nu c_\rho d_\sigma \ . 
\labbel{defe4} \ee 
Eq.(\ref{MVSchid}) is known as the Schouten identity \cite{MV}; 
by squaring it, one gets a huge polynomial, of fifth-order 
in the scalar products of all the vectors. Due to Eq.(\ref{MVSchid}), 
that polynomial vanishes in $d=4$ dimensions (and, {\it a fortiori} for 
any integer dimension $d\le4$); note however that the polynomial does not 
vanish identically for any arbitrary value of the dimension; as 
Eq.(\ref{MVSchid}) is valid only when $d\le4$, for $d > 4$ the polynomial 
is not bound to take a vanishing value. \par 
As an extension (or rather a simplification) of Eq.(\ref{MVSchid}), 
consider now the quantity 
\be \epsilon(a,b) =  \epsilon_{\mu\nu}a_\mu b_\nu \ , \labbel{defe2} \ee 
where  $ \epsilon_{\mu\nu} $ is the Levi-Civita tensor with two indices 
(defined of course by $ \epsilon_{12} = - \epsilon_{21} =1, $ 
$ \epsilon_{11} = \epsilon_{22} = 0 $), and $ a_\mu,b_\nu $ are a couple 
of two-dimensional vectors. By squaring it, Eq.(\ref{defe2}) gives at once 
\be \epsilon^2(a,b) = a^2 b^2 - (a\cdot b)^2 \ , \labbel{defe2sq} \ee 
where $ a^2, b^2 $ are the squared moduli of the vectors $ a_\mu,b_\nu $ 
and $ (a\cdot b) $ their scalar product. \\ 
So far, all the quantities introduced in Eq.s(\ref{defe2},\ref{defe2sq}) 
are in $ d=2 $ dimensions. If the dimension $ d $ takes the value of 
any (non-vanishing) integer less than $ 2 $ ({\it i.e.} if $ d=1$), 
the {\it r.h.s.} of Eq.(\ref{defe2}) vanishes, and so does the 
{\it r.h.s.} of Eq.(\ref{defe2sq}) as well. 
At this point we define the Schouten Polynomial $ P_2(d;a,b) $ as 
\be P_2(d;a,b) = a^2 b^2 - (a\cdot b)^2 \ , \labbel{defP2} \ee 
where the {\it r.h.s.} is formally the same {\it r.h.s.} of 
Eq.(\ref{defe2sq}), but the two vectors $ a_\mu,b_\nu $ are assumed to 
be $d$-dimensional vectors, with continuous $ d $. To emphasize 
that point, we have written $ d $ within the arguments of the 
Schouten Polynomial, even if $ d $ does not appear explicitly in the 
{\it r.h.s.} of Eq.(\ref{defP2}). By the very definition, 
at integer non vanishing dimension $d<2$ ({\it i.e.} at $ d=1$), 
$ P_2(d;a,b) $ vanishes, 
\be P_2(1;a,b) = 0 \ , \labbel{defP2(1)} \ee 
as can be also verified by an absolutely trivial explicit calculation. \par 
Following the elementary procedure leading to Eq.(\ref{defP2}), given 
in $d=3$ dimensions any triplet of vectors $ a_\mu,b_\nu,c_\rho $ we 
consider 
\be \epsilon(a,b,c) =  \epsilon_{\mu\nu\rho}a_\mu b_\nu c_\rho \ , 
\labbel{defe3} \ee
where $ \epsilon_{\mu\nu\rho} $ is the Levi-Civita tensor with three 
indices (defined as usual by $ \epsilon_{123} = 1 $ {\it etc.}) and 
then we evaluate its square 
\be \epsilon^2(a,b,c) = a^2 b^2 c^2 - a^2(b\cdot c)^2 - b^2(a\cdot c)^2 
   - c^2(a\cdot b)^2 + 2(a\cdot b)(b\cdot c)(a\cdot c) \ . 
\labbel{defe3sq} \ee 
We then define the Schouten Polynomial $ P_3(d;a,b,c) $ as 
\be P_3(d;a,b,c) = a^2 b^2 c^2 - a^2(b\cdot c)^2 - b^2(a\cdot c)^2 
   - c^2(a\cdot b)^2 + 2(a\cdot b)(b\cdot c)(a\cdot c) \ . 
\labbel{defP3} \ee 
where again the {\it r.h.s.} is formally the same as in Eq.(\ref{defe3sq}), 
but the three vectors $ a_\mu,b_\nu,c_\rho $ are assumed to 
be $d$-dimensional vectors, with continuous $ d $. By construction, 
$ P_3(d;a,b,c) $ vanishes at $d=1$ and at $d=2$, 
\begin{align} P_3(1;a,b,c) &= 0 \ , \nonumber\\ 
              P_3(2;a,b,c) &= 0 \ . \labbel{defP3(1,2)} 
\end{align} 
Needless to say, the procedure can be immediately iterated to any higher 
dimension, generating Schouten polynomials involving four vectors and 
vanishing in $ d=1,2,3 $ dimensions, or involving five vectors and 
vanishing in $ d=1,2,3,4 $ dimensions, corresponding, up to a constant 
numerical factor, to the square of Eq.(\ref{MVSchid}), {\it etc.}. \par 
As it is apparent from the previous discussion, the Schouten polynomials 
generated by a given set of vectors are nothing but the Gram determinants 
of the corresponding vectors; we prefer to refer to them as Schouten 
polynomials to emphasize that they vanish in any integer dimension $ d $ 
less than the number of the vectors. 
\par 
In the actual physical applications, as one is interested mainly in the 
$ d \to 4 $ limit of Feynman graph amplitudes, one can reach $ d=4 $ 
starting from a different value of $ d $ and then moving to $ d=4 $ by 
means of the Tarasov's shifting relations\cite{Tarasov1996}. As the shift 
relates values of $ d $ differing by two units, the $ d=1 $ Schouten 
identities, easily established for any amplitude in which at least two 
vectors occur, are of no use. The next simplest identities are at $ d=2 $ 
and occur with any amplitude involving at least three vectors. That is 
the case of the two-loop self-mass graph (the sunrise), which we will 
study in this paper in the arbitrary mass case. 
\section {The Schouten Identities for the Sunrise graph} 
\labbel{SchIds} 

We discuss in this Section the use of the Schouten polynomial 
$ P_3(d;a,b,c) $ in the case of the sunrise, the two-loop self-mass 
graph of Fig.(1). \par 
The external momentum is $ p $ and the internal masses are $ m_1, m_2, m_3 $. 
We use the Euclidean metric, so that $ p^2 $ is positive when spacelike; 
sometimes we will use also $ s = W^2 = -p^2 $, so that the sunrise amplitudes 
develop an imaginary part when $ \sqrt{s} = W  > (m_1+m_2+m_3) $, the 
threshold of the Feynman graph. 
We write the propagators as 
\begin{align} D_1 &= q_1^2 + m_1^2 \ , \nonumber\\ 
              D_2 &= q_2^2 + m_2^2 \ , \nonumber\\ 
              D_3 &= (p-q_1-q_2)^2 + m_3^2 \ , \labbel{defDi} 
\end{align} 
and define the loop integration measure, in agreement with previous works, as:
\be \intQ = \frac{1}{C(d)} \int \frac{d^d q}{(2 \pi)^{d-2}}\,,\ee
with 
\be C(d) = (4 \pi)^{(4-d)/2} \Gamma\left( 3 - \frac{d}{2}\right) \ , 
\labbel{defCd} \ee 
so that 
\be C(2) = 4\pi \quad \mbox{and}\quad C(4) = 1 \ . \labbel{C(2,4)} \ee 
With that definition the Tadpole $ T(d,m) $ reads
\be T(d;m) = \intQ \ \frac{1}{q^2+m^2} 
           =  \frac{m^{d-2}}{(d-2)(d-4)} \ . \labbel{defTad} \ee 
In this paper we will use the ``double" tadpoles 
\be T(d;m_1,m_2) =  \intq \frac{1}{D_1 D_2} \ , \labbel{defT} \ee 
together with the similarly defined $ T(d;m_1,m_3), T(d;m_2,m_3) $, 
and the four amplitudes 
\begin{align} S(d;p^2) &= \intq \frac{1}{D_1 D_2 D_3} \ , \nonumber\\ 
 S_1(d;p^2) &= - \frac{d}{dm_1^2} S(d;p^2) 
                        = \intq \frac{1}{D_1^2 D_2 D_3} \ , \nonumber\\ 
 S_2(d;p^2) &= - \frac{d}{dm_2^2} S(d;p^2) 
                        = \intq \frac{1}{D_1 D_2^2 D_3} \ , \nonumber\\ 
 S_3(d;p^2) &= - \frac{d}{dm_3^2} S(d;p^2) 
                        = \intq \frac{1}{D_1 D_2 D_3^2} \ . 
\labbel{defSi} \end{align} 
All those amplitudes depend on the three masses $ m_1, m_2, m_3 $, even if 
the masses are not written explicitly in the arguments for simplicity. 
The four amplitudes are equal, when multiplied by an overall constant factor 
$ (2\pi)^4 $, to the four M.I.s used in \cite{Laporta1998}. $ S(d;p^2) $, 
in particular, is the full scalar amplitude already referred to 
previously. Those amplitudes were chosen in \cite{Laporta1998} as 
M.I.s for the sunrise problem, and in the following they will be 
sometimes referred to as the ``conventional'' M.I.s . 
\par 
We can now introduce the {\it Schouten amplitudes} defined, for 
arbitrary $ d $, as 
\be Z(d;n_1,n_2,n_3,p^2) = \intq \frac{P_3(d;p,q_1,q_2)} 
           { D_1^{n_1}D_2^{n_2}D_3^{n_3} } \labbel{defZs} \ , \ee 
where the $ n_i $ are positive integer numbers and $P_3(d;p,q_1,q_2)$ 
is the Schouten polynomial defined in Eq.(\ref{defP3}). The convergence of 
the integrals, for a given value of $ d $, depends of course on the 
powers $ n_i $, as the Schouten polynomial in the numerator contributes 
always with four powers of the loop momenta $ q_1 $ and $ q_2 $. \par 
We are interested here in the $ d=2 $ case. If the Schouten amplitude is
convergent at $ d=2 $, due to the second of Eq.s(\ref{defP3(1,2)}),
it is also vanishing at $ d=2 $, {\it i.e.} $ Z(2;n_1,n_2,n_3,p^2) = 0 $.
Note that in the massive case all the integrals we are considering
are {\it i.r.} finite, therefore the divergences can only be of {\it u.v.}
nature. \par
As one can express any sunrise Feynman amplitude in terms of a 
valid set of M.I.s, we will write in the following a few 
Schouten amplitudes in terms of the ``conventional'' M.I.s given in 
Eq.s(\ref{defSi}). A few explicit results are now listed: 
\begin{align} Z_1(d;p^2) &= Z(d;1,2,2) \nonumber\\ 
       &= \frac{(d-1)}{12} \left[ -(d-2)p^2 + (d-3)( -2m_1^2+m_2^2+m_3^2 ) 
                                        ) \right] S(d;p^2) \nonumber\\ 
       &- \frac{(d-1)}{6} (p^2+m_1^2) \ m_1^2 S_1(d;p^2)   \nonumber\\
       &+ \frac{(d-1)}{12} ( p^2 -3m_1^2 +m_2^2 +3m_3^2 ) 
                                        \ m_2^2 S_2(d;p^2) \nonumber\\ 
       &+ \frac{(d-1)}{12} ( p^2 -3m_1^2 +3m_2^2 +m_3^2 ) 
                                        \ m_3^2 S_3(d;p^2) \nonumber\\
       &+ \frac{(d-1)(d-2)}{24}  \left[ T(d;m_1,m_2) + T(d;m_1,m_3) 
                                  - 2\,T(d;m_2,m_3) \right] \ , 
\labbel{defZd1} \end{align}
\begin{align}  Z_2(d;p^2) &= Z(d;2,1,2,p^2) \nonumber\\ 
       &=  \frac{(d-1)}{12} \left[ -(d-2)p^2 + (d-3)(m_1^2 -2m_2^2 +m_3^2 ) 
                                           \right] S(d;p^2) \nonumber\\ 
       &+ \frac{(d-1)}{12} ( p^2 +m_1^2 -3m_2^2 +3m_3^2 ) 
                                         \ m_1^2 S_1(d,p^2) \nonumber\\
       &- \frac{(d-1)}{6} (p^2+m_2^2) \ m_2^2 S_2(d;p^2) \nonumber\\ 
       &+ \frac{(d-1)}{12} ( p^2 +3m_1^2 - 3m_2^2 +m_3^2 ) 
                                       \ m_3^2 S_3(d;p^2) \nonumber\\ 
       &+ \frac{(d-1)(d-2)}{24} \left[ T(d;m_1,m_2) - 2\,T(d;m_1,m_3) 
                                    + T(d;m_2,m_3) \right] \ , 
\labbel{defZd2} \end{align} 
\begin{align} Z_3(d;p^2) &= Z(d;2,2,1,p^2) \nonumber\\ 
       &= \frac{(d-1)}{12} \left[ -(d-2)p^2 + (d-3)(m_1^2+m_2^2-2m_3^2) 
                           \right] S(d;p^2) \ ,  \nonumber\\ 
       &+\frac{(d-1)}{12} ( p^2 +m_1^2 +3m_2^2 -3m_3^2 ) 
                                       \ m_1^2 S_1(d;p^2) \nonumber\\ 
       &+\frac{(d-1)}{12} ( p^2 +3m_1^2 +m_2^2 -3m_3^2 ) 
                                       \ m_2^2 S_2(d;p^2) \nonumber\\ 
       &-\frac{(d-1)}{6}   (p^2+m_3^2) \ m_3^2 S_3(d;p^2) \nonumber\\ 
       &+ \frac{(d-1)(d-2)}{24} \left[ - 2\,T(d;m_1,m_2) + T(d;m_1,m_3) 
                                            + T(d;m_2,m_3) \right] \ , 
\labbel{defZd3} \end{align} 
\begin{align} Z(d;2,2,2,p^2) &= - \frac{(d-1)(d-2)}{4} \nonumber\\ 
           & \times \left[ (d-3) S(d;p^2) 
          + m_1^2 S_1(d;p^2) + m_2^2 S_2(d;p^2) + m_3^2 S_3(d;p^2) 
            \right] \ . 
\labbel{defZd222} \end{align}
Some comments are in order. 
Elementary power counting arguments give $ N=2(n_1+n_2+n_3) $ powers of 
the integration momenta in the denominator (independently of $d$) and, 
in $d=2$ dimensions, all together eight powers in the numerator 
(see Eq.(\ref{defZs}) for the definition of the integrals), 
so that the minimum value of $ N $ necessary to guarantee the convergence 
is $ N=10 $. In the case of $ Z(d;2,2,2,p^2) $ of Eq.(\ref{defZd222}) 
$ N = 12 $, more than the minimum required value $ N=10 $; 
therefore the integrals in the loop momenta $q_1, q_2$ do converge, 
so that the vanishing of $ P_3(d;p,q_1,q_2) $ in the numerator at $ d=2 $ 
(and, as a matter of fact at $ d=1 $ as well) does imply the vanishing 
of the whole amplitude. The explicit result, Eq.(\ref{defZd222}), shows 
indeed that the amplitude vanishes at $ d=1 $ and $d=2$, but that is due 
to an overall factor $ (d-1)(d-2) $, so that Eq.(\ref{defZd222}) 
does not give any useful information. This pattern -- the vanishing at 
$ d=2 $ of the amplitudes with $ P_3(d;p,q_1,q_2) $ in the numerator and 
$ N> 10 $ due to the appearance of an overall factor $(d-2)$ -- is of 
general nature, and showed up in all the cases which we were able to check 
(needless to say, the algebraic complications increase quickly with the 
powers of the denominators). 
\par 
The $ Z_i(d;p^2) $, Eq.s(\ref{defZd1},\ref{defZd2},\ref{defZd3}), 
are more interesting; in this case, $ N=10 $, which is the minimum 
value needed to guarantee convergence in $ d=2 $ dimensions, so that 
those amplitudes are expected to vanish at $ d=2 $ (and therefore 
also at $d=1$) as a consequence of the vanishing of $ P_3(d;p,q_1,q_2) $ 
at $ d=1, d=2 $, see Eq.s(\ref{defP3(1,2)}). The vanishing at $ d=1 $ is 
trivially given by the overall factor $(d-1)$ (in $ d=1 $ the minimum 
value of $ N $ to guarantee convergence is $ N=8 $, while in the 
integrals we are now considering $ N=10 $), but the vanishing at $ d=2 $ 
is totally non trivial, providing new (and so far not known) relations 
between the four conventional M.I.s $ S(d;p^2), S_i(d;p^2) $ at $ d=2 $. 
\par 
Any of the three amplitudes $ Z_i(d;p^2) $ can obviously be obtained from 
the others by a suitable permutation of the three masses, as immediately 
seen from their explicit expression. When summing the three relations, 
one obtains 
\be Z_1(d;p^2) + Z_2(d;p^2) + Z_3(d;p^2) = 
           - \frac{(d-1)(d-2)}{4} p^2 \ S(d;p^2) \ , 
\labbel{sumZs} \ee 
showing that at $ d=2 $ they are not independent from each other; 
indeed, if one takes as input $ Z_2(2;p^2) = 0 $ and $ Z_3(2;p^2) = 0 $, 
the previous equation gives $ Z_1(2;p^2) = 0 $, showing that the condition 
$ Z_1(2;p^2) =0 $ depends on the other two. When written explicitly, the 
vanishing of $ Z_2(2;p^2) = 0 $ and $ Z_3(2;p^2) = 0 $ reads 
\begin{align}  Z_2(2;p^2) &= - \frac{1}{12} (m_1^2 -2m_2^2 +m_3^2 ) 
                                           S(2;p^2) \nonumber\\ 
       &+ \frac{1}{12} ( p^2 +m_1^2 -3m_2^2 +3m_3^2 ) 
                                         \ m_1^2 S_1(2,p^2) \nonumber\\
       &- \frac{1}{6} (p^2+m_2^2) \ m_2^2 S_2(2;p^2) \nonumber\\ 
       &+ \frac{1}{12} ( p^2 +3m_1^2 - 3m_2^2 +m_3^2 ) 
                                       \ m_3^2 S_3(2;p^2) \nonumber\\ 
       &+ \frac{1}{96}\ln\frac{m_2^2}{m_1m_3} 
       \nonumber\\ &= 0 \ , 
\labbel{defZ22} \end{align} 
\begin{align} Z_3(2;p^2) &= -\frac{1}{12} (m_1^2+m_2^2-2m_3^2) 
                              S(2;p^2) \ ,  \nonumber\\ 
       &+\frac{1}{12} ( p^2 +m_1^2 +3m_2^2 -3m_3^2 ) 
                                       \ m_1^2 S_1(2;p^2) \nonumber\\ 
       &+\frac{1}{12} ( p^2 +3m_1^2 +m_2^2 -3m_3^2 ) 
                                       \ m_2^2 S_2(2;p^2) \nonumber\\ 
       &-\frac{1}{6}   (p^2+m_3^2) \ m_3^2 S_3(2;p^2) \nonumber\\ 
       &+ \frac{1}{96}\ln\frac{m_3^2}{m_1m_2} 
       \nonumber\\ &= 0 \ .
\labbel{defZ32} \end{align} 
The validity of identities Eq.s(\ref{defZ22},\ref{defZ32}) in $d=2$ has been 
checked with SecDec~\cite{SecDec2}.
By using the above relations, which  hold identically in $ p^2, m_1^2, 
m_2^2, m_3^2 $, one can express two of the conventional M.I.s in terms 
of the other two, showing that, at $d=2$, there are in fact only two 
independent M.I.s. As can be seen from Eq.s(\ref{defZ22},\ref{defZ32}), 
the relations between the M.I.s are not trivial (in particular, none of 
the M.I.s vanishes at $ d=2 $; according to the definition 
Eq.(\ref{defSi}) for space-like $ p $ they are all positive definite). 
\section {A New Set of Master Integrals} \labbel{NSMI} 

We have seen in the previous Section that the ``conventional" M.I.s 
in $ d=2 $ dimensions satisfy two independent conditions, 
written explicitly in Eq.s(\ref{defZ22},\ref{defZ32}), so that 
two of them can be expressed as a combination of the other two, 
which can be taken as independent. On the other hand, 
it is known that in the equal mass limit the Sunrise has two independent 
M.I.s (in any dimension, including $d=2$) so that no other 
independent conditions can exist. It can therefore be convenient to 
introduce a new set of M.I.s, formed by two ``conventional" M.I.s , say 
$ S(d;p^2), S_1(d;p^2) $ of Eq.(\ref{defSi}), and two Schouten amplitudes, 
say $ Z_2(d;p^2), Z_3(d;p^2) $ of Eq.s(\ref{defZd2},\ref{defZd3}). 
The advantage of the choice is that two conditions at $ d=2 $ take the 
simple form $ Z_2(2;p^2) = 0, \ Z_3(2;p^2) = 0. $ The actual choice of 
the new M.I.s satisfying the above criteria is of course not unique 
(a fully equivalent set could be for instance $ S(d;p^2), S_2(d;p^2), 
Z_1(d;p^2), Z_2(d;p^2) $ {\it etc}.). \par 
In the new basis of M.I.s, the two discarded conventional M.I.s are 
expressed as 
\begin{align} 
 P(p^2,m_1,m_2,m_3)\,&m_2^2\,S_2(d;p^2) = \nonumber\\ 
  &\ \Big\{ (m_1^2-m_2^2) \left[ (d-3)(m_1^2+m_2^2-m_3^2)-p^2 \right] 
       - (d-2)p^2 (p^2+m_3^2) \Big\} S(d;p^2) \nonumber\\
       &+ P(p^2,m_2,m_1,m_3)\,m_1^2\,S_1(d;p^2) \nonumber\\ 
       &- \frac{8}{(d-1)} \left(p^2 + m_3^2\right) Z_2(d;p^2) \nonumber\\ 
       &- \frac{4}{(d-1)} \left(p^2 + 3m_1^2 - 3m_2^2 + m_3^2\right) 
                                                   Z_3(d;p^2) \nonumber\\ 
       &- (d-2) \left(m_1^2 - m_2^2\right)  T(d;m_1,m_2)      \nonumber\\ 
       &- \frac{(d-2)}{2} \,\left(p^2 - m_1^2 + m_2^2 + m_3^2\right) 
                                                 T(d;m_1,m_3) \nonumber\\ 
       &+ \frac{(d-2)}{2} \,\left(p^2 + m_1^2 - m_2^2 + m_3^2\right) 
                                                  T(d;m_2,m_3)\ , 
\labbel{S2Z} \end{align} 
\begin{align} 
 P(p^2,m_1,m_2,m_3)\,&m_3^2\,S_3(d;p^2) = \nonumber\\ 
  &\ \Big\{ (m_1^2-m_3^2) \left[(d-3)(m_1^2-m_2^2+m_3^2)-p^2\right] 
         - (d-2)p^2 (p^2+m_2^2) \Big\} S(d;p^2) \nonumber\\ 
       &+ P(p^2,m_3,m_1,m_2)\,m_1^2\,S_1(d;p^2) \nonumber\\ 
       &- \frac{4}{(d-1)} \left(p^2 + 3m_1^2 + m_2^2 - 3 m_3^2\right) 
                                                   Z_2(d;p^2) \nonumber\\ 
       &- \frac{8}{(d-1)} \left(p^2 + m_2^2\right) Z_3(d;p^2) \nonumber\\ 
       &- \frac{(d-2)}{2} \,\left(p^2 - m_1^2 + m_2^2 + m_3^2\right) 
                                                 T(d;m_1,m_2) \nonumber\\ 
       &- (d-2) \left(m_1^2 - m_3^2\right)  T(d;m_1,m_3)      \nonumber\\ 
       &+ \frac{(d-2)}{2} \,\left(p^2 + m_1^2 + m_2^2 - m_3^2\right) 
                                                  T(d;m_2,m_3)\ , 
\labbel{S3Z} \end{align} 
where $ P(p^2,m_1,m_2,m_3) $ is the polynomial 
\begin{align} P(p^2,m_1,m_2,m_3) &= p^4 + 2(m_2^2+m_3^2-m_1^2)p^2 \nonumber\\ 
  &- 3m_1^4+m_2^4+m_3^4+2m_1^2m_2^2+2m_1^2m_3^2-2m_2^2m_3^2 \ . 
\labbel{defP} \end{align} 
Note that $ P(p^2,m_1,m_2,m_3) $, which is symmetric in the last two 
arguments, 
\be P(p^2,m_1,m_2,m_3) = P(p^2,m_1,m_3,m_2) \ , \labbel{symP} \ee 
occurs with different arguments in different places. \par 
By substituting the above expressions in the differential equations 
for the conventional M.I.s as given, for instance, in 
Ref.\cite{Laporta1998}, one obtains the new equations 
\begin{align} 
 P(p^2,m_1,m_2,m_3) \,p^2 \frac{d}{dp^2} \,S(d;p^2) &= (p^2 + m_1^2) 
    \Big[ (p^2 - m_1^2 + m_2^2 + m_3^2) \nonumber\\ 
  &+ (d-2)( p^2 + m_1^2 - m_2^2 - m_3^2) \Big]\,S(d;p^2)\nonumber\\
  &- Q(p^2,m_1,m_2,m_3)\,m_1^2\,S_1(d;p^2) \nonumber\\
  &+\frac{4}{(d-1)} (3 p^2 + 3 m_1^2 + m_2^2 - m_3^2)\, Z_2(d;p^2)\nonumber\\ 
  &+\frac{4}{(d-1)} (3 p^2 + 3 m_1^2 - m_2^2 + m_3^2)\, Z_3(d;p^2)\nonumber\\
  &+\frac{(d-2)}{2}(p^2 + m_1^2 - m_2^2 + m_3^2)\,T(d;m_1,m_2)\nonumber\\
  &+\frac{(d-2)}{2}(p^2 + m_1^2 + m_2^2 - m_3^2)\,T(d;m_1,m_3)\nonumber\\
  &-(d-2)\,(p^2 + m_1^2)\,T(d;m_2,m_3) \ , 
\labbel{eq:diff1} \end{align}
\begin{align} 
 D(p^2,m_1,m_2,m_3)&P(p^2,m_1,m_2,m_3) \,p^2 \frac{d}{dp^2} \,S_1(d;p^2) = 
 \nonumber \\
  &\Big[ \frac{(d-2)^2}{2}\,\left(p^2 + m_1^2 - m_2^2 - m_3^2\right) 
                       \,P_{10}^{(2)}(p^2,m_1,m_2,m_3) \nonumber\\  
  &  - (d-2)\,P_{10}^{(1)}(p^2,m_1,m_2,m_3) 
   - P_{10}^{(0)}(p^2,m_3,m_1,m_2)\, \Big] S(d;p^2) \nonumber \\
  &+ \left[ \frac{(d-2)}{2}\,P_{11}^{(1)}(p^2,m_1,m_2,m_3)
   - P_{11}^{(0)}(p^2,m_1,m_2,m_3) \right] \,S_1(d;p^2)\nonumber \\ 
  &+\frac{4(d-3)}{(d-1)}\Big[ P_{12}^{(0)}(p^2,m_1,m_2,m_3)\,Z_2(d;p^2) 
   + P_{12}^{(0)}(p^2,m_1,m_3,m_2)\,Z_3(d;p^2) \Big]\nonumber\\
  &+\frac{(d-2)}{4}\,\left[\frac{(d-2)}{m_1^2}\,P_{14}^{(2)}(p^2,m_1,m_2,m_3) 
   - 2\,P_{14}^{(1)}(p^2,m_1,m_2,m_3)\right]\,T(d;m_1,m_2)\nonumber\\
  &+\frac{(d-2)}{4}\,\left[\frac{(d-2)}{m_1^2}\,P_{14}^{(2)}(p^2,m_1,m_3,m_2) 
   - 2\,P_{14}^{(1)}(p^2,m_1,m_3,m_2)\,\right]\,T(d;m_1,m_3)\nonumber\\
  &-\frac{(d-2)}{2}\,\Big[(d-2)\,P_{10}^{(2)}(p^2,m_1,m_2,m_3) \nonumber \\
  &\qquad \qquad  -  \left( P_{14}^{(1)}(p^2,m_1,m_2,m_3) 
   + P_{14}^{(1)}(p^2,m_1,m_3,m_2) \right)\,\Big]\,T(d;m_2,m_3)  \ , 
\labbel{eq:diff2} \end{align} 

\begin{align} 
   P(p^2,m_1,m_2,m_3) \,p^2 \frac{d}{dp^2} \,&Z_2(d;p^2) = p^2 
   \,\frac{(d-1)(d-2)}{8}\,\Big[ 2 \left(  m_1^2 -  m_2^2\right) 
   \left( p^2 +  m_1^2 +  m_2^2 -  m_3^2 \right)\nonumber\\
  &+ (d-2)\left(p^2 +  m_1^2 -  m_2^2 -  m_3^2\right) 
   \left(p^2 +  m_1^2 -  m_2^2 +  m_3^2\right)\Big]\, S(d;p^2) \nonumber\\
  &- p^2\,\frac{(d-1)(d-2)}{4}\,P(p^2,m_2,m_1,m_3)\,m_1^2\,S_1(d;p^2) 
                                                               \nonumber\\ 
  &+ \frac{(d-2)}{2}\,P_{22}(p^2,m_1,m_2,m_3)\,Z_2(d;p^2)\nonumber\\
  &+ p^2\,(d-2)\,\left(p^2 + 3  m_1^2 - 3  m_2^2 +  m_3^2\right) 
                                                   \,Z_3(d;p^2)\nonumber\\
  &+ p^2\,\frac{(d-1)(d-2)^2}{4}\,\left( m_1^2 -  m_2^2\right) 
                                                 \,T(d;m_1,m_2)\nonumber\\
  &+ p^2\,\frac{(d-1)(d-2)^2}{8}\,\left(p^2 -  m_1^2 +  m_2^2 
                                  +  m_3^2\right)\,T(d;m_1,m_3)\nonumber\\
  &- p^2\,\frac{(d-1)(d-2)^2}{8}\,\left(p^2 +  m_1^2 -  m_2^2 
                                  +  m_3^2\right)\,T(d;m_2,m_3) \ , 
\labbel{eq:diff3} \end{align} 
\begin{align} 
   P(p^2,m_1,m_2,m_3) \,p^2 \frac{d}{dp^2} \,&Z_3(d;p^2) = p^2 
   \,\frac{(d-1)(d-2)}{8}\,\Big[ 2 \left(  m_1^2 -  m_3^2\right) 
   \left( p^2 +  m_1^2 -  m_2^2 +  m_3^2 \right)\nonumber\\
  &+ (d-2)\left(p^2 +  m_1^2 -  m_2^2 -  m_3^2\right) \left(p^2 +  m_1^2 
   +  m_2^2 -  m_3^2\right)\Big]\, S(d;p^2) \nonumber\\
  &- p^2\,\frac{(d-1)(d-2)}{4}\,P(p^2,m_3,m_1,m_2)\,m_1^2\,S_1(d;p^2) 
                                                               \nonumber\\
  &+ p^2\,(d-2)\,\left(p^2 + 3  m_1^2 +  m_2^2 - 3  m_3^2\right) 
                                                   \,Z_2(d;p^2)\nonumber\\
  &+ \frac{(d-2)}{2}\,P_{22}(p^2,m_1,m_3,m_2)\,Z_3(d;p^2)\nonumber\\
  &+ p^2\,\frac{(d-1)(d-2)^2}{8}\,\left(p^2 -  m_1^2 +  m_2^2 
                                   +  m_3^2\right)\,T(d;m_1,m_2)\nonumber\\
  &+ p^2\,\frac{(d-1)(d-2)^2}{4}\,\left( m_1^2 -  m_3^2\right) 
                                                  \,T(d;m_1,m_3)\nonumber\\
  &- p^2\,\frac{(d-1)(d-2)^2}{8}\,\left(p^2 +  m_1^2 +  m_2^2 
                                  -  m_3^2\right)\,T(d;m_2,m_3) \ . 
\labbel{eq:diff4} \end{align} 
In the above equations, 
\begin{align} D(p^2,m_1,m_2,m_3) = 
    &(p^2+(m_1+m_2+m_3)^2)(p^2+(m_1-m_2+m_3)^2) \nonumber\\ 
    &(p^2+(m_1+m_2-m_3)^2)(p^2+(m_1-m_2-m_3)^2) 
\labbel{defD} \end{align} 
is the product of all the threshold and pseudo-threshold factors already 
present in \cite{Laporta1998}, 
\begin{align}
 Q(p^2,m_1,m_2,m_3) =  &- (m_1+m_2+m_3) (m_1-m_2+m_3) 
 (m_1+m_2-m_3) (m_1-m_2-m_3) \nonumber\\
       &+2\, p^2  \left( m_1^2+m_2^2+m_3^2 \right)
        + 3\,p^4\,, \labbel{defQ}
\end{align}
while $ P(p^2,m_1,m_2,m_3) $ is the 
polynomial previously defined in Eq.(\ref{defP}). 
Finally the $ P_{ij}^{(n)}(p^2,m_1,m_2,m_3) $ are also 
polynomials depending on $ p^2 $ and the masses; their explicit (and 
sometimes lengthy expression) is given in Appendix~\ref{App:Poly1}. 
Note that a same polynomial can occur in different equations with a 
different permutation of the masses within its arguments. \newline

We want to stress here an important aspect of the last two equations,
Eq.(\ref{eq:diff3},\ref{eq:diff4}), namely the presence of an overall 
factor $ (d-2) $ in the {\it r.h.s.}, which plays an important role 
in the expansion in powers of $(d-2)$ discussed in the next Subsection. 
\subsection {The expansion of the Equations around $d=2$ } \labbel{EExd2} 
Let us start off by expanding all M.I.s in powers of $(d-2)$ around $ d=2 $, 
\begin{align} 
   S(d;p^2)   &= S(2;p^2) + (d-2)S^{(1)}(2,p^2) + ... \nonumber\\ 
   S_1(d;p^2) &= S_1(2;p^2) + (d-2)S_1^{(1)}(2,p^2) + ... \nonumber\\ 
   Z_2(d;p^2) &= Z_2(2;p^2) + (d-2)Z_2^{(1)}(2,p^2) + ... \nonumber\\ 
   Z_3(d;p^2) &= Z_3(2;p^2) + (d-2)Z_3^{(1)}(2,p^2) + ... 
\labbel{expd-2} \end{align} 
Due to the overall factor $(d-2)$ in the {\it r.h.s.}, 
at $0$th order in $(d-2)$ the differential equations 
Eq.s(\ref{eq:diff3},\ref{eq:diff4}) become 
\begin{align}
 \frac{d}{dp^2} \,Z_2(2;p^2) &= 0 \nonumber\\
 \frac{d}{dp^2} \,Z_3(2;p^2) &= 0 \ , \labbel{eqZd2} 
\end{align} 
showing that $ Z_2(2;p^2), Z_3(2;p^2) $ must be constants. But we know 
from Eq.s(\ref{defZ22},\ref{defZ32}) the actual value of that 
constant (the two functions vanish 
identically, $  Z_2(2;p^2) = 0,\  Z_3(2;p^2) = 0\ $), so that at 
$0$th order in $(d-2)$ the differential equations Eq.(\ref{eq:diff1}), 
Eq.(\ref{eq:diff2}) for $ S(2;p^2), S_1(2;p^2) $ become 
\begin{align}
   P(p^2,m_1,m_2,m_3) &\,p^2 \frac{d}{dp^2} \,S(2;p^2) =
       \left(p^2 + m_1^2 \right) \, \left(p^2 - m_1^2 + m_2^2 + m_3^2 
                                     \right) \,S(2;p^2) \nonumber\\ 
  & - Q(p^2,m_1,m_2,m_3) m_1^2 \,S_1(2;p^2) \nonumber \\
  &+ \frac{1}{8} \Big[ \left(p^2 + m_1^2\right) \ln{\frac{m_1^2}{m_2m_3}} 
   + \left(m_2^2 - m_3^2 \right) \ln{\frac{m_3}{m_2}} \Big] \ , 
\labbel{eqSd2} \end{align} 
\begin{align} 
   D(p^2,m_1,m_2,m_3)&P(p^2,m_1,m_2,m_3) \,p^2 \frac{d}{dp^2} \,S_1(2;p^2) =
   \nonumber\\
  &-  P_{10}^{(0)}(p^2,m_1,m_2,m_3)  \,S(2,p^2) 
   -  P_{11}^{(0)}(p^2,m_1,m_2,m_3)  \,S_1(2,p^2) \nonumber\\ 
  &- \frac{1}{8}  \Bigl[
      P_{14}^{(1)}(p^2,m_1,m_2,m_3)\ln{\frac{m_1}{m_3}} 
   +  P_{14}^{(1)}(p^2,m_1,m_3,m_2)\ln{\frac{m_1}{m_2}} \nonumber\\
  &       -  \frac{p^2}{m_1^2} \, P^2(p^2,m_1,m_2,m_3) \Bigr]  \ , 
\labbel{eqS1d2} \end{align} 
completely decoupled, obviously, from the (trivial) equations for 
$ Z_2(2;p^2), Z_3(2;p^2). $ (See the previous Section and 
Appendix~\ref{App:Poly1} for the explicit expression of the polynomials.) 
\par 
Going now one order higher in the expansion, one finds that 
the first-order terms in $(d-2)$ of the $ Z_i(d;p^2) $ 
satisfy the equations 
\begin{align} 
   P(p^2,m_1,m_2,m_3) \,\frac{d}{dp^2} \,Z_2^{(1)}(2;p^2) &= \frac{1}{4} 
     (m_1^2 - m_2^2) (p^2 + m_1^2 + m_2^2 - m_3^2) \,S(2;p^2) 
                                                              \nonumber\\
  & - \frac{1}{4} P(p^2,m_2,m_1,m_3)\, m_1^2\, S_1(2;p^2) \nonumber\\
  &+ \frac{1}{32} \Big[ (p^2 + m_3^2) \ln\frac{m_1}{m_2} 
                        + (m_1^2 - m_2^2)\ln\frac{m_1m_2}{m^2_3} \Big] \ , 
\labbel{eqZ21d2} \end{align}
\begin{align}
   P(p^2,m_1,m_2,m_3) \,\frac{d}{dp^2} \,Z_3^{(1)}(2;p^2) &= \frac{ 1}{4} 
     (m_1^2 - m_3^2) (p^2 + m_1^2 - m_2^2 + m_3^2)\,S(2;p^2) 
                                                             \nonumber \\
  &-  \frac{ 1}{4}  P(p^2,m_3,m_1,m_2)\,m_1^2\,S_1(2;p^2) \nonumber \\
  &+ \frac{1}{32} \Big[ (p^2 + m_2^2) \ln\frac{m_1}{m_3} 
                        + (m_1^2 - m_3^2)\ln\frac{m_1m_3}{m^2_2} \Big] \ . 
\labbel{eqZ31d2} \end{align} 
It is to be noted that $ Z_2^{(1)}(2;p^2), Z_3^{(1)}(2;p^2) $ do not appear 
in the {\it r.h.s.} of Eq.s(\ref{eqZ21d2},\ref{eqZ31d2}), which contains 
only $ S(2;p^2) $ and $ S_1(2;p^2), $ to be considered known once 
Eq.s(\ref{eqSd2},\ref{eqS1d2}) for the $0th$ orders in $(d-2)$ have been 
solved. Eq.s(\ref{eqZ21d2},\ref{eqZ31d2}), indeed, are absolutely trivial 
when considered as differential equations, as they contain only 
the derivatives of $ Z_2^{(1)}(2;p^2), Z_3^{(1)}(2;p^2), $ and 
can therefore be solved by a simple quadrature. 
\par 
Knowing $ Z_2^{(1)}(2;p^2), Z_3^{(1)}(2;p^2) $, one can move to the 
differential equations for $ S^{(1)}(2;p^2), S_1^{(1)}(2;p^2) $ (which we 
don't write here for the sake of brevity); they involve 
$ Z_2^{(1)}(2;p^2), Z_3^{(1)}(2;p^2) $ as known inhomogeneous terms, 
and form again a closed set of two differential equations, 
decoupled from the equations for the other two M.I.s, as at 
$0th$ order in $(d-2)$. \par 
Thanks to the overall factor $(d-2)$ in the {\it r.h.s.} of 
Eq.s(\ref{eq:diff3},\ref{eq:diff4}), the pattern 
-- a quadrature for 
$ Z_2^{(k)}(2;p^2), Z_3^{(k)}(2;p^2) $ and a closed set of two 
differential equations for $ S^{(k)}(2;p^2), S_1^{(k)}(2;p^2) $ -- 
is completely general, and can be iterated, at least in principle, 
up to any required order $k$ in $(d-2)$. 

\section{Second-order Differential Equation for $S(d;p^2)$}
\labbel{IIordDE}
We go back now to the system of differential equations 
Eq.s(\ref{eq:diff1},\ref{eq:diff2}), for obtaining a second-order differential 
equation for $ S(d;p^2). $ 
We can use Eq.(\ref{eq:diff1}) in order to express
$S_1(d;p^2)$ in function of $S(d;p^2)$ and of its derivative, $dS(d;p^2)/dp^2$.
By substituting this expression into Eq.(\ref{eq:diff2}) 
we can then derive a second-order differential equation 
for $S(d;p^2)$ only, which however
still contains $Z_2(d;p^2)$ and $Z_3(d;p^2)$ in the inhomogeneous part:
\begin{align}
A_1(p^2,m_1,m_2,m_3)\, &\left( {d \over dp^2} \right)^2 S(d;p^2) 
+ \Big[A_2^{(0)}(p^2,m_1,m_2,m_3) 
+ (d-2)\,A_2^{(1)}(p^2,m_1,m_2,m_3)\Big]\, 
  {d \over dp^2} S(d;p^2)\nonumber \\ 
  &+ (d-3)\Big[A_3^{(0)}(p^2,m_1,m_2,m_3)
     + (d-2)\,A_3^{(1)}(p^2,m_1,m_2,m_3)\Big]\, S(d;p^2) \nonumber \\ 
&+\frac{(d-3)}{(d-1)} \, \Big[ A_4(p^2,m_1,m_2,m_3)\,Z_2(d;p^2) 
   + A_4(p^2,m_1,m_3,m_2)\,Z_3(d;p^2) \,\Big] \nonumber \\ 
&+ (d-2) \,\Big[\, A_5^{(1)}(p^2,m_1,m_2,m_3) 
  + (d-2)\,A_5^{(2)}(p^2,m_1,m_2,m_3) \,\Big] T(d;m_1,m_2) \nonumber \\ 
&+ (d-2) \,\Big[\, A_5^{(1)}(p^2,m_1,m_3,m_2) 
  + (d-2)\,A_5^{(2)}(p^2,m_1,m_3,m_2) \,\Big] T(d;m_1,m_3) \nonumber \\
&+ (d-2) \,\Big[\, A_5^{(1)}(p^2,m_2,m_3,m_1) 
  + (d-2)\,A_5^{(2)}(p^2,m_2,m_3,m_1) \,\Big] T(d;m_2,m_3) \nonumber \\
&= 0\,, \labbel{eq:IIordd}
\end{align}
where 
$ A_1(p^2,m_1,m_2,m_3) = 
 p^2\,D(p^2,m_1^2,m_2^2,m_3^2)\,P(p^2,m_1,m_2,m_3)\,,
$
with $D(p^2,m_1^2,m_2^2,m_3^2)$ and \newline
$P(p^2,m_1,m_2,m_3)$ being the usual polynomials
defined by Eq.s(\ref{defP},\ref{defD}). The $A_j^{(n)}(p^2,m_1,m_2,m_3)$ are
also polynomials which depend on the three 
masses and on $p^2$, but \textsl{do not} depend on the dimensions $d$.
Their explicit expressions, as usual quite lengthy, 
can be found in Appendix~\ref{App:Poly2}. 
\newline

The equation above is exact in $d$ but contains, 
besides $ S(d;p^2) $ and its derivatives, also 
$Z_2(d;p^2)$ and $Z_3(d;p^2)$ as inhomogeneous terms. 
Nevertheless, recalling once more that $ Z_2(2;p^2) = Z_3(2;p^2) = 0 $, 
we can expand Eq.(\ref{eq:IIordd}) in powers of $(d-2)$ and obtain at 
leading order in $ (d-2) $ a second-order differential equation 
for $S(2;p^2)$ \textsl{only}:
\begin{align}
A_1(p^2,m_1,m_2,m_3)\, &\left( {d \over dp^2} \right)^2 \, S(2;p^2) 
+ A_2^{(0)}(p^2,m_1,m_2,m_3)  \left({d \over dp^2} \right)\, S(2;p^2)
\nonumber \\ &
- A_3^{(0)}(p^2,m_1,m_2,m_3) \, S(2;p^2) 
+ \frac{1}{4}\,\Bigg[ A_5^{(2)}(p^2,m_1,m_2,m_3) \nonumber \\ 
&+ A_5^{(2)}(p^2,m_1,m_3,m_2) 
+ A_5^{(2)}(p^2,m_2,m_3,m_1) \nonumber\\
&+A_5^{(1)}(p^2,m_1,m_2,m_3) \,\ln{\left( \frac{m_1}{m_3} \right)} 
+ A_5^{(1)}(p^2,m_1,m_3,m_2) \,\ln{\left(\frac{m_1}{m_2}\right)} \Bigg] = 0\,,
\labbel{eq:IIord2}
\end{align} 
where we made use of the relation Eq.(\ref{relApol}) of 
Appendix~\ref{App:Poly2}.
We compared Eq.(\ref{eq:IIord2}) with the second-order differential equation 
derived in~\cite{Weinzierl2012}, finding perfect agreement. 
Eq.(\ref{eq:IIord2}) has been solved in reference~\cite{Weinzierl2013}
in terms of one-dimensional integrals over elliptic integrals. \par
Upon inserting the result in Eq.(\ref{eqSd2}) one can obtain
$ S_1(2;p^2) $ in terms of $ S(2;p^2) $ and $ dS(2;p^2)/dp^2 $. 
Inserting then $ S(2;p^2) $ and $ S_1(2;p^2) $ in Eq.s(\ref{eqZ21d2}, 
\ref{eqZ31d2}), one obtains by quadrature the first-order 
terms, $ Z_2^{(1)}(2;p^2) $ and $ Z_3^{(1)}(2;p^2) $, 
of the expansion in $(d-2)$ of $Z_2(d;p^2)$ and $Z_3(d;p^2)$. \par 

Having these results on hand, we can now consider the  
first-order in $ (d-2) $ of the 
Eq.(\ref{eq:IIordd}), which is now a second-order differential equation for 
$ S^{(1)}(2;p^2) $ only, with known inhomogeneous terms 
(namely $S(2;p^2)$, $Z_2^{(1)}(2;p^2)$ and $Z_3^{(1)}(2;p^2)$). 
Proceeding in this way, at least in principle, 
the whole procedure can be iterated up to any required order 
in $(d-2)$. 

\section {Shifting relations from $d=2$ to $d=4$ dimensions} 
\labbel{Shift} 
In the previous Sections we have shown how to use the Schouten identities 
for writing the differential equations for the M.I.s of the massive 
sunrise at $ d=2 $ dimensions in block form, and outlined the procedure 
for obtaining iteratively all the coefficients of the expansion in $(d-2)$ 
of the four M.I.s starting from a second-order differential equation for 
$ S(2;p^2) $, the leading term of the expansion. \par 
The physically interesting case corresponds however to the expansion 
of the M.I.s for $ d\approx4 $; we have therefore 
to convert the information given by the expansion at $ d\approx2 $ in useful 
information at $ d\approx4 $. \par 
As it is well known, quite in general one can relate any Feynman integral 
evaluated in $d$ dimensions to the very same integral evaluated in $(d-2)$ 
dimensions by means of the Tarasov's shifting relation~\cite{Tarasov1996}. 
This dimensional shift is achieved by acting on the Feynman integral 
with a suitable combination of derivatives with respect to the internal
masses.
In the case of the ``conventional " M.I.s of the sunrise graph, as defined 
in Eq.({\ref{defSi}), the shifting relations read: 
\begin{align}
S(d-2;p^2) &= \frac{2^2}{(d-6)}\, \Delta\, S(d;p^2)\,,\nonumber \\
S_i(d-2;p^2) &= \frac{2^2}{(d-6)}\, \Delta\, S_i(d;p^2)\,, 
\qquad i=1,2,3\,,\labbel{eq:dirshift}
\end{align}
where the differential operator $\Delta$ takes the form:
\be 
\Delta =  \frac{\partial}{\partial m_1^2} \frac{\partial}{\partial m_2^2} 
  + \frac{\partial}{\partial m_1^2} \frac{\partial}{\partial m_3^2} 
  + \frac{\partial}{\partial m_2^2} \frac{\partial}{\partial m_3^2}\,. 
  \labbel{eq:diffop} 
\ee 
Carrying out the derivatives in the integral representation for the 
four M.I.s of Eq.(\ref{defSi}), one obtains a combination of integrals which 
are still related to the sunrise graph. They can be expressed in terms of the 
full set of M.I.s in $d$ dimensions (by full set we mean the four M.I.s 
and the tadpoles); one obtains in that way a set of four equations which 
explicitly relate the four M.I.s of the sunrise graph evaluated in $(d-2)$ 
dimensions to suitable combinations of the same integrals (and of the 
tadpoles) evaluated in $d$ dimensions. In that {\it direct} form 
the shifting relations would be of no practical use in 
our case, as they might give the M.I.s at $ (d-2) \approx 2 $ in terms of 
those (less known) at $ d \approx 4 $. \par 
It is however straightforward to invert the system and, in this way, to 
obtain the {\it inverse} shifting relations, expressing the four M.I.s 
in $d+2 \approx 4 $ dimensions in function of those 
in $ d\approx 2 $ dimensions. In addition, we can also use 
Eq.s(\ref{S2Z},\ref{S3Z}) for expressing $S_2(d;p^2)$ and $S_3(d;p^2)$,
in terms of $S(d;p^2)$, $S_1(d;p^2)$ and $Z_2(d;p^2)$, $Z_3(d;p^2)$. 
As a result one arrives at expressing any of the four ``conventional" 
M.I.s $ S(d+2,p^2) $, $ S_i(d+2;p^2) $, $i=1,2,3$, as a linear combination 
(whose coefficients depend -- and in a non trivial way -- on $ d $ and 
the kinematical variables of the problem) of the ``new" M.I.s 
$S(d;p^2)$, $S_1(d;p^2)$, $Z_2(d;p^2)$ and $Z_3(d;p^2)$ (and the 
tadpoles). Indicating for simplicity the four ``conventional" M.I.s with 
$ M_i(d) $ and with $ N_j(d) $ the four ``new" M.I.s and the tadpoles, and 
ignoring for ease of notation all the kinematical variables, the 
{\it inverse} shifting relations can be written as 
\be M_i(d+2) = \sum_j C_{i,j}(d) N_j(d) \ . \labbel{Tarinv} \ee 
Given a relation of the form 
$$ F(d+2) = G(d) \ , $$ 
by expanding around $ d=2 $ one has, quite in general 
\begin{align} 
 F(d+2) &= \sum_{n=r}^p (d-2)^n\ F^{(n)}(4) \ , \nonumber\\ 
   G(d) &= \sum_{n=r}^p (d-2)^n\ G^{(n)}(2) \ , \nonumber 
\end{align} 
where $ r $, the first value of the summation index, can be negative 
(as it is the case in a Laurent expansion), so that 
$$ F^{(n)}(4) = G^{(n)}(2) \ . $$ 
In the case of the {\it inverse} shift Eq.(\ref{Tarinv}), one has that 
the coefficients of the expansion of the ``conventional" M.I.s in 
$(d-4)$ for $ d \approx 4 $ are completely determined by those of the 
expansion in $(d-2) $ for $ d \approx 2 $ of the ``new" M.I.s,  
discussed in the previous Sections, and of the tadpoles (expanding around 
$ d=2 $ the two sides of Eq.(\ref{Tarinv}) requires also the expansion of 
the coefficients $ C_{i,j}(d) $, but that is not a problem once the 
{\it inverse} shift has been written down explicitly). \par 
The explicit formulas of the {\it direct} or {\it inverse} shifting 
relations are easily obtained but very lengthy and we decided
not to include them entirely here for the sake of brevity. 
For what follows, it is sufficient to discuss only the general features 
of one of the {\it inverse} shifting relations, 
namely the relation expressing $S(d+2;p^2)$ in terms of 
$S(d;p^2)$, $S_1(d;p^2)$ and $Z_2(d;p^2)$, $Z_3(d;p^2)$. 
Keeping for simplicity only the leading term of the expansion in $(d-2)$ 
of the coefficients we find: 
\begin{align}
S(d+2;p^2) &= \left[ \,C(p^2,m_1,m_2,m_3) 
+ \mathcal{O}\left( d-2\right) \, \right] S(d;p^2) \nonumber \\
&+  \left[ \,C_1(p^2,m_1,m_2,m_3) 
+ \mathcal{O}\left( d-2\right) \, \right] S_1(d;p^2) \nonumber \\
&+  \left[ \,\frac{1}{d-2}\ C_2(p^2,m_1,m_2,m_3) 
+ \mathcal{O}\left( 1\right) \, \right] Z_2(d;p^2) \nonumber \\
&+  \left[ \,\frac{1}{d-2}\ C_3(p^2,m_1,m_2,m_3) 
+ \mathcal{O}\left( 1\right) \, \right] Z_3(d;p^2) \nonumber \\
&+  \left[ \,C_4^{(0)}(p^2,m_1,m_2,m_3) 
+ \mathcal{O}\left( d-2\right) \, \right] T(d;m_1,m_2) \nonumber \\
&+  \left[ \,C_5^{(0)}(p^2,m_1,m_2,m_3)
+ \mathcal{O}\left( d-2\right) \, \right] T(d;m_1,m_2) \nonumber \\
&+  \left[ \,C_6^{(0)}(p^2,m_1,m_2,m_3) 
+ \mathcal{O}\left( d-2\right) \, \right] T(d;m_1,m_2)\,.
 \labbel{eq:tarinv}
\end{align}
In the formula above the $C(p^2,m_1,m_2,m_3)$, $C_i(p^2,m_1,m_2,m_3)$, 
are ratios of suitable polynomials which, as usual, depend on
$p^2$ and on the three masses but, most importantly, 
\textsl{do not} depend on the dimensions $d$.  
The explicit expressions for 
$C(p^2,m_1,m_2,m_3)$, $C_1(p^2,m_1,m_2,m_3)$,
$C_2(p^2,m_1,m_2,m_3)$ and $C_3(p^2,m_1,m_2,m_3)$,
which will also be used in the following, 
can be found in Appendix~\ref{App:Tar}, Eq.s(\ref{defC}-\ref{defC3}).
Note anyway that: 
$$C_3(p^2,m_1,m_2,m_3) = C_2(p^2,m_1,m_3,m_2)\,.$$ 
By writing the expansion of $ S(d+2;p^2) $ at $ d \approx 2 $ as 
\be  S(d+2;p^2) = \sum_n S^{(n)}(4;p^2) (d-2)^n\,, \labbel{expd4} \ee 
and then expanding Eq.(\ref{eq:tarinv}) at $ d \approx 2 $, 
one recovers the expression of the coefficients $S^{(n)}(4;p^2)$ 
in terms of the coefficients of the expansion of the four M.I.s 
and the tadpoles in $(d-2)$.
\par 
A few observations are in order. 
Eq.(\ref{eq:tarinv}) 
exhibits an explicit pole in $ 1/(d-2) $ 
only in the coefficients of 
$ Z_2(d;p^2) $ and $ Z_3(d;p^2) $; recalling once more that at $ d=2 $ 
both $Z_2(2;p^2)$ and $ Z_3(2;p^2) $ are identically zero, 
see Eq.s(\ref{defZ22},\ref{defZ32}), it is clear that 
these poles will not generate any 
singularity of $ S(d;p^2) $ as $d\to2$. 
On the other hand, the tadpoles in the {\it r.h.s.} of Eq.(\ref{eq:tarinv}) 
do generate polar singularities of $ S(d+2;p^2) $; recalling 
Eq.s(\ref{defTad},\ref{defT}) and by using the lengthy explicit form of the 
coefficients (which we did not write for brevity) one finds immediately 
\begin{align} 
   S^{(-2)}(4;p^2) &= -\frac{(m_1^2 + m_2^2 + m_3^2)}{8} \,, \nonumber\\ 
      S^{(-1)}(4;p^2) &=  
    \frac{1}{32} \Big[ p^2 + 6 \left( m_1^2 + m_2^2 + m_3^2\right) \Big]\nonumber  \\
      &-\frac{1}{8} \Big[ m_1^2\,\ln{(m_1^2)} + m_2^2\,\ln{(m_2^2)} + m_3^2\,\ln{(m_3^2)}  \Big],
\end{align} 
formulas already known for a long time in the literature~\cite{Laporta1998}.
\par 
As a second observation, let us look at the zeroth-order term 
$ S^{(0)}(4;p^2) $ of $ S(d;p^2) $ in $(d-4)$, {\it i.e.\ } the zeroth-order 
term in $ (d-2) $ of Eq.(\ref{eq:tarinv}). We have already commented the 
apparent polar singularity 
$1/(d-2)$ in the coefficients of $Z_2(d;p^2)$ and $Z_3(d;p^2)$, actually 
absent because  $Z_2(2;p^2)$ and $ Z_3(2;p^2) $ are both vanishing. 
But due to the presence of the $ 1/(d-2) $ 
polar factor, in order to recover the zeroth-order term 
$ S^{(0)}(4;p^2) $, one needs, besides $ S(2;p^2), S_1(2;p^2), $ also the 
first-order of the corresponding expansion of $Z_2(d;p^2) $ and 
$ Z_3(d;p^2) $, namely $ Z_2^{(1)}(2;p^2) $ and $ Z_3^{(1)}(2;p^2) $ 
-- obtained, in our approach, from the systematic expansion of 
the differential equations, see Eq.s(\ref{eqZ21d2},\ref{eqZ31d2}) or 
Section~\ref{IIordDE}. \par 
The complete expression of $S^{(0)}(4;p^2)$, which is rather cumbersome, 
is given by Eq.(\ref{eq:shiftS4}) of  Appendix~\ref{App:Tar}. 
The corresponding formulas for the other three M.I.s, 
i.e. the $S_i(d+2;p^2)$, can be obtained directly from the authors. 
\section {The imaginary parts of the Master Integrals. } 
\labbel{imMI} 
In this Section, which is somewhat pedagogical, 
we discuss the relationship between the imaginary parts of 
the M.I.s at $ d=2 $ and $ d=4 $ dimensions, as a simple but explicit 
example of functions exhibiting the properties described in the previous 
sections. \par 
At $ d=2 $ the Cutkosky-Veltman rule\cite{Cutkosky1960,Veltman1963}  
gives for $ S(d;p^2) $, as defined by the first of Eq.s(\ref{defSi}),
\be \frac{1}{\pi}ImS(2;-W^2) = N_2\int_{b_2}^{b_3} db 
    \ \frac{1}{\sqrt{R_4(b;b_1,b_2,b_3,b_4)}} \ , \labbel{Im2S} \ee 
where the following notations were introduced: 
\begin{align} 
 & N_2 = 1/2 \nonumber\\ 
 & p^2 = - W^2, \hspace{1cm} W \ge m_1+m_2+m_3 \ , \nonumber\\ 
 & (m_2-m_3)^2 = b_1 \le (m_2+m_3)^2 = b_2 \le (W-m_1)^2 = b_3 
                     \le (W+m_1)^2 = b_4 \ , \nonumber\\ 
 & R_4(b;b_1,b_2,b_3,b_4) = (b-b_1)(b-b_2)(b_3-b)(b_4-b) \ . 
\labbel{defbiR4} \end{align} 
We have the relation 
\be R_4(b;b_1,b_2,b_3,b_4) = R_2(b,m_2^2,m_3^2)\ R_2(W^2,b,m_1^2) \ , 
                                                  \labbel{R4R2R2} \ee 
where 
\be R_2(a,b,c) = a^2+b^2+c^2-2ab-2ac-2bc \ , \labbel{defR2} \ee 
is the familiar invariant form appearing in the two-body phase space, 
showing that the system of the three particles, whose masses enter in the 
definition of $ R_4(b;b_1,b_2,b_3,b_4) $, can be considered as the merging 
of a two-body system of total energy $ \sqrt{b} $ and masses $ m_2,m_3 $ 
with a two-body system of total energy $ W $ and masses 
$ \sqrt{b}, m_1 $ . \\ 
According to Eq.s(\ref{defSi}), for i=1,2,3 
\be \frac{1}{\pi}ImS_i(2;-W^2) = 
  - \frac{d}{dm_i^2} \left( \frac{1}{\pi}ImS(2;-W^2) \right) \ ; 
\labbel{Im2Si} \ee 
the integral representation Eq.(\ref{Im2S}), however, is of no 
use for obtaining $ ImS_i(2;-W^2) $ through a direct differentiation 
(due to the appearance of end point singularities). 
It is more convenient to use Eq.(\ref{KtoI0a}) of the Appendix~\ref{App:im}, so that 
Eq.(\ref{Im2S}) becomes 
\be \frac{1}{\pi}ImS(2;-W^2) = N_2 
          \frac{2}{ \sqrt{(b_4-b_2)(b_3-b_1)} } K(w^2) \ , \labbel{Im2SK} \ee 
where $ K(w^2) $ is the complete elliptic integral of the first kind, 
Eq.(\ref{defK}), and 
\begin{align} 
  w^2 &= \frac{(b_4-b_1)(b_3-b_1)}{(b_4-b_2)(b_3-b_1)} \nonumber\\ 
      &= \frac{ (W+m_1+m_2-m_3)(W+m_1-m_2+m_3) 
                (W-m_1+m_2-m_3)(W-m_1-m_2+m_3) } 
              { (W+m_1+m_2+m_3)(W+m_1-m_2-m_3) 
                (W-m_1+m_2-m_3)(W-m_1-m_2+m_3) } \ , \nonumber\\ 
  (b_4-b_2)&(b_3-b_1) = (W+m_1+m_2+m_3)(W+m_1-m_2-m_3) \nonumber\\ 
  &\ \ \ \ \ \ \ \ \ \ \ \times(W-m_1+m_2-m_3)(W-m_1-m_2+m_3) \ . 
                                                    \labbel{Im2SKa} 
\end{align} 
Let us observe, in passing, that, even if  $ ImS(2,-W^2) $ (and, more 
generally $ S(d;p^2) $ as well) is obviously symmetric in the three 
masses $ m_1,m_2,m_3 $, the symmetry is not explicitly shown by the 
integral representation Eq.(\ref{Im2S}), while the manifest symmetry 
is restored in Eq.s(\ref{Im2SK},\ref{Im2SKa}). \\ 
One can now use Eq.(\ref{Im2SK}) and Eq.(\ref{derK}) to carry out the 
derivative with respect to the masses $ m_i^2 $ in Eq.(\ref{Im2Si}); the 
result reads 
\begin{align}
 \frac{1}{\pi}ImS_1(2;-W^2) &= N_2 
                \frac{1}{2m_1^2\sqrt{(b_3-b_1)(b_4-b_2)}} 
                \frac{1}{(b_3-b_2)(b_4-b_1)}\, \nonumber \\ 
   &\times \left[ 4 m_1 (m_1 m_3^2 + m_1 m_2^2 - m_1^3 + 2 m_2 m_3 W 
                 + m_1 W^2)\,K(w^2)  \right. \nonumber \\ 
   &\left.- P( - W^2,m_1,m_2,m_3) \,E(w^2)  \right]\,, \labbel{Im2S1K} 
\end{align}
\begin{align} 
 \frac{1}{\pi}ImS_2(2;-W^2) &= N_2 
                \frac{1}{2m_2^2\sqrt{(b_3-b_1)(b_4-b_2)}} 
                \frac{1}{(b_3-b_2)(b_4-b_1)}\, \nonumber \\ 
   &\times \left[ 4 m_2 (m_2 m_3^2 + m_2 m_1^2 - m_2^3 + 2 m_1 m_3 W 
                 + m_2 W^2)\,K(w^2)  \right. \nonumber \\ 
   &\left.- P( - W^2,m_2,m_1,m_3) \,E(w^2)  \right]\,, \labbel{Im2S2K} 
\end{align} 
\begin{align}
 \frac{1}{\pi}ImS_3(2;-W^2) &= N_2 
                \frac{1}{2m_3^2\sqrt{(b_3-b_1)(b_4-b_2)}} 
                \frac{1}{(b_3-b_2)(b_4-b_1)}\, \nonumber \\ 
   &\times \left[ 4 m_3 (m_3 m_1^2 + m_3 m_2^2 - m_3^3 + 2 m_2 m_1 W 
                 + m_3 W^2)\,K(w^2)  \right. \nonumber \\ 
   &\left.- P( - W^2,m_3,m_2,m_1) \,E(w^2)  \right]\,, \labbel{Im2S3K}
\end{align} 
where $ w^2 $ is given by the first of Eq.s(\ref{Im2SKa}), 
$ P(p^2,m_1,m_2,m_3) $ is the polynomial already introduced in 
Eq.(\ref{defP}), symmetric in the last two arguments, and $ E(w^2) $ 
is the complete elliptic integral of the second kind, see Eq.(\ref{defE}). 
\par 
Eq.s(\ref{Im2SK},\ref{Im2S1K},\ref{Im2S2K},\ref{Im2S3K}) express the 
four quantities $ ImS(2;-W^2), ImS_i(2,-W^2), i=1,2,3 $ in terms of just 
two functions, the elliptic integrals $ K(w^2), E(w^2) $; therefore, the 
four imaginary parts cannot be all linearly independent. It is indeed 
easy to check that they satisfy the two equations 
\begin{align}  
&- \frac{1}{12} (m_1^2 -2m_2^2 +m_3^2 ) \ Im S(2;-W^2) 
 + \frac{1}{12} ( -W^2 +m_1^2 -3m_2^2 +3m_3^2 ) 
                                         \ m_1^2\ ImS_1(2,-W^2) \nonumber\\
&- \frac{1}{6} (-W^2+m_2^2) \ m_2^2 \ ImS_2(2;-W^2) 
 + \frac{1}{12} ( -W^2 +3m_1^2 - 3m_2^2 +m_3^2 ) 
                           \ m_3^2 \ ImS_3(2;-W^2) = 0 \ , \nonumber 
\end{align} 
\begin{align} 
&- \frac{1}{12} (m_1^2+m_2^2-2m_3^2) \ ImS(2;-W^2) 
 + \frac{1}{12} ( -W^2 +m_1^2 +3m_2^2 -3m_3^2 ) 
                                       \ m_1^2\ Im S_1(2;-W^2) \nonumber\\ 
&+ \frac{1}{12} ( -W^2 +3m_1^2 +m_2^2 -3m_3^2 ) 
                                       \ m_2^2\ ImS_2(2;-W^2) 
 - \frac{1}{6} ( -W^2+m_3^2) \ m_3^2\ ImS_3(2;-W^2) = 0 \ , \nonumber 
\end{align} 
which are nothing but the imaginary parts of $ Z_2(2;-W^2), Z_3(2;-W^2) $, 
Eq.s(\ref{defZ22},\ref{defZ32}). \par 
As a further comment on the imaginary parts at $ d=2 $, let us observe 
that they take a finite value at threshold, {\it i.e.} in the 
$ W \to (m_1+m_2+m_3) $ limit. In that limit, indeed, 
$ b_3 \to b_2 = (m_2+m_3)^2 $, and one finds 
$$ \int_{b_2}^{b_3} \frac{db}{\sqrt{R_4(b;b_1,b_2,b_3,b_4)}} 
   \to \frac{1}{ \sqrt{(b_2-b_1)(b_4-b_2)} } 
   \int_{b_2}^{b_3} \frac{db}{ \sqrt{(b-b_2)(b_3-b)} } 
   = \frac{\pi}{ \sqrt{(b_2-b_1)(b_4-b_2)} } \ , 
$$ 
so that 
\be  \frac{1}{\pi}ImS(2;-W^2) \xrightarrow[W \to (m_1+m_2+m_3) ]{} 
         \frac{N_2}{4\sqrt{m_1m_2m_3(m_1+m_2+m_3)} } \ . 
\labbel{Im2Sthr} \ee 
The extension to the value at threshold of the $ ImS_i(2,-W^2) $ is 
similar, even if requiring one more term in the expansion due to the 
presence of the denominator $ 1/(b_3-b_2) $ in their definitions, 
Eq.s(\ref{Im2S1K},\ref{Im2S2K},\ref{Im2S3K}). The threshold values are 
\begin{align} 
  \frac{1}{\pi}ImS_1(2;-W^2) &\xrightarrow[W \to (m_1+m_2+m_3) ]{} 
                                             \nonumber\\ 
    & \frac{N_2}{32}\left( - \frac{3}{m_1} + \frac{1}{m_2} + \frac{1}{m_3} 
    - \frac{1}{m_1+m_2+m_3} \right) 
         \frac{1}{\sqrt{m_1m_2m_3(m_1+m_2+m_3)} } \ , \nonumber\\ 
  \frac{1}{\pi}ImS_2(2;-W^2) &\xrightarrow[W \to (m_1+m_2+m_3) ]{} 
                                             \nonumber\\ 
    & \frac{N_2}{32}\left( + \frac{1}{m_1} - \frac{3}{m_2} + \frac{1}{m_3} 
    - \frac{1}{m_1+m_2+m_3} \right) 
         \frac{1}{\sqrt{m_1m_2m_3(m_1+m_2+m_3)} } \ , \nonumber\\ 
  \frac{1}{\pi}ImS_3(2;-W^2) &\xrightarrow[W \to (m_1+m_2+m_3) ]{} 
                                             \nonumber\\ 
    & \frac{N_2}{32}\left( + \frac{1}{m_1} + \frac{1}{m_2} - \frac{3}{m_3} 
    - \frac{1}{m_1+m_2+m_3} \right) 
         \frac{1}{\sqrt{m_1m_2m_3(m_1+m_2+m_3)} } \ . \labbel{Im2Sithr} 
\end{align} 

At $ d=4 $ the imaginary part of $ S(d;p^2) $, by using the same notation 
as in Eq.(\ref{Im2S}), is given by 
\be \frac{1}{\pi}ImS(4;-W^2) = N_4\int_{b_2}^{b_3} db 
    \ \frac{1}{b} \sqrt{R_4(b;b_1,b_2,b_3,b_4)} \ . \labbel{Im4S} \ee 
with 
\be N_4 = \frac{1}{8 W^2} \ . \labbel{defN4} \ee 
At variance with the $ d=2 $ case, the $ ImS_i(4,-W^2) $ can be 
obtained at once by differentiating with respect to the masses the 
previous integral representation for $ ImS(4;-W^2) $. The result can be 
most conveniently expressed in terms of the four (independent) 
integrals $ I(-1,W), I(0,W), I(1,W), I(2,W), $ defined 
(see Eq.(\ref{defIn}) and the Appendix for more details and the relation 
to the standard complete elliptic integrals) through 
\be I(n,W) = \int_{b_2}^{b_3} db\ b^n 
         \frac{1}{ \sqrt{R_4(b;b_1,b_2,b_3,b_4)} } \ . \labbel{defInn} \ee 
An explicit calculation gives 
\begin{align} 
     \frac{1}{\pi}ImS(4;-W^2) = N_4 \Bigl[&\ b_1 b_2 b_3 b_4 \ I(-1,W) 
                                                            \nonumber\\ 
   &- \frac{3}{4}(b_2 b_3 b_4 + b_1 b_3 b_4 + b_1 b_2 b_4 + b_1 b_2 b_3) 
                                                \ I(0,W) \nonumber\\ 
   &+ \frac{1}{2} ( b_3 b_4 + b_2 b_4 + b_2 b_3 + b_1 b_4 
                            + b_1 b_3 + b_1 b_2)\ I(1,W) \nonumber\\ 
   &- \frac{1}{4} (b_1 + b_2 + b_3 + b_4 )\ I(2,W)\ \Bigr] \labbel{Im4SI} 
\end{align} 
\begin{align} 
      \frac{1}{\pi}ImS_1(4;-W^2) = N_4 \Bigl[& 
      b_1 b_2 \bigl( - (b_4-b_3)W + (b_4 + b_3 )m_1 \bigr) 
                                               \ I(-1,W) \nonumber\\ 
   &+ \bigl( (b_2+b_1)(b_4-b_3)W 
    - (b_2 b_4 + b_2 b_3 + b_1 b_4 + b_1 b_3 + 2 b_1 b_2)m_1 \bigr) 
                                                 \ I(0,W) \nonumber\\ 
   &+ \bigl( (b_4-b_3)W + (b_4 + b_3 + 2b_2 + 2b_1)m_1 \bigr) 
                                                 \ I(1,W) \nonumber\\ 
   &- 2m_1 \                             I(2,W)\ \Bigr] \labbel{Im4S1I} 
\end{align} 
\begin{align} 
      \frac{1}{\pi}ImS_2(4;-W^2) = N_4 \Bigl[& 
      b_3 b_4 \bigl( -( b_2-b_1)m_3 + (b_2+b_1)m_2 \bigr) 
                                               \ I(-1,W) \nonumber\\ 
   &+ \bigl( - (2 b_3 b_4 + b_2 b_4 + b_2 b_3 + b_1 b_4 + b_1 b_3) m_2 
             + (b_2-b_1)(b_4+b_3) m_3 \bigr)    \ I(0,W) \nonumber\\ 
   &+ \bigl( (2 b_4 + 2 b_3 + b_2 + b_1) m_2 - (b_2-b_1) m_3 \bigr) 
                                                 \ I(1,W) \nonumber\\ 
   &- 2m_2 \                             I(2,W)\ \Bigr] \labbel{Im4S2I} 
\end{align} 
\begin{align} 
      \frac{1}{\pi}ImS_3(4;-W^2) = N_4 \Bigl[& 
      b_3 b_4 \bigl( -( b_2-b_1)m_2 + (b_2+b_1)m_3 \bigr) 
                                               \ I(-1,W) \nonumber\\ 
   &+ \bigl( - (2 b_3 b_4 + b_2 b_4 + b_2 b_3 + b_1 b_4 + b_1 b_3) m_3 
             + (b_2-b_1)(b_4+b_3) m_2 \bigr)    \ I(0,W) \nonumber\\ 
   &+ \bigl( (2 b_4 + 2 b_3 + b_2 + b_1) m_3 - (b_2-b_1) m_2 \bigr) 
                                                 \ I(1,W) \nonumber\\ 
   &- 2m_3 \                             I(2,W)\ \Bigr] \labbel{Im4S3I} 
\end{align} 
Again at variance with the $ d=2 $ case, the four imaginary parts are 
now combinations of four independent elliptic integrals, and therefore
all independent of each other. \par 
\medskip 
Having recalled the main features of the imaginary parts of the M.I.s 
at $ d=2 $ and $ d=4 $ dimensions, we can look at the way the Tarasov's 
shifting relations work in their case. \par 
Let us start from the ``direct" shift 
expressing the imaginary parts at $ d=2 $ in terms of those at $ d=4 $. 
The $ d \to 4 $ limit of the shifting relations is trivial, even if the 
relevant formulas are as usual rather lengthy. 
Keeping only the imaginary parts of the master integrals one finds 
for the M.I. $ S(2,p^2) $, with $ -p^2 = W^2 \ge (m_1+m_2+m_3)^2 $ 
\begin{align} \frac{1}{\pi}ImS(2,-W^2) &= 
        \tilde{A}(W,m_1,m_2,m_3) \frac{1}{\pi}ImS(4,-W^2) \nonumber\\ 
    &+\tilde{B}(W,m_1,m_2,m_3) m_1\frac{1}{\pi}ImS_1(4,-W^2) \nonumber\\ 
    &+\tilde{B}(W,m_2,m_3,m_1) m_2\frac{1}{\pi}ImS_2(4,-W^2) \nonumber\\ 
    &+\tilde{B}(W,m_3,m_1,m_2) m_3\frac{1}{\pi}ImS_3(4,-W^2) \ , 
\labbel{Im2SIm4} 
\end{align} 
where 
\begin{align} 
  \tilde{A}(W,m_1,m_2,m_3) &= 
          {A}(W,m_1,m_2,m_3) + {A}(W,m_1,-m_2,m_3) \nonumber\\ 
      &+ {A}(W,m_1,m_2,-m_3) + {A}(W,m_1,-m_2,-m_3) \ , \nonumber\\ 
  \tilde{B}(W,m_1,m_2,m_3) &= 
          {B}(W,m_1,m_2,m_3) + {B}(W,m_1,-m_2,m_3) \nonumber\\ 
      &+ {B}(W,m_1,m_2,-m_3) + {B}(W,m_1,-m_2,-m_3) \ , \nonumber\\ 
 {A}(W,m_1,m_2,m_3) &= \frac{1}{2m_1m_2m_3} 
          \ \frac{m_1+m_2+m_3}{W^2-(m_1+m_2+m_3)^2} \ , \nonumber\\ 
 {B}(W,m_1,m_2,m_3) &= \frac{1}{2}(2m_1+m_2+m_3){A}(W,m_1,m_2,m_3) \ . 
\labbel{Im2SIm4a} 
\end{align} 
Eq.(\ref{Im2SIm4}) is relatively simple, and, when substituting in it the 
explicit values of $ ImS(4,-W^2) $ and $ ImS_i(4,-W^2) $, as given by 
Eq.s(\ref{Im4SI}--\ref{Im4S3I}), Eq.(\ref{Im2S}) is recovered. 
The same happens for $ ImS_i(2,-W^2),$ $ i=1,2,3 \ $ as well. \par 
Conversely, one can look at the inverse formulas, giving the imaginary 
parts at $ d+2 \to 4 $ in terms of the imaginary parts at $ d \to 2 $. 
For $ ImS(4,-W^2) $, taking only the imaginary part at $ d=2 $ 
of Eq.(\ref{eq:tarinv}),  one obtains:
\begin{align} 
   \frac{1}{\pi}ImS(4;-W^2) &= C(-W^2,m_1,m_2,m_3) \frac{1}{\pi}ImS(2;-W^2) 
                                                 \nonumber\\ 
  &+ C_1(-W^2,m_1,m_2,m_3) \frac{1}{\pi}ImS_1(2;-W^2) \nonumber\\ 
  &+ C_2(-W^2,m_1,m_2,m_3) \frac{1}{\pi}ImZ_2^{(1)}(2;-W^2) \nonumber\\ 
  &+ C_3(-W^2,m_1,m_2,m_3) \frac{1}{\pi}ImZ_3^{(1)}(2;-W^2) \ , 
                                                       \labbel{Im4SIm2a} 
\end{align} 
where the $ C(-W^2,m_1,m_2,m_3), C_i(-W^2,m_1,m_2,m_3) $ 
have been defined in the previous section, and their explicit expressions
can be found in Eq.s(\ref{defC}-\ref{defC3}), 
$ ImS(2;-W^2), ImS_1(2;-W^2) $ are the imaginary parts of 
the corresponding Master Integrals at $ d=2 $, while 
$ ImZ_2^{(1)}(2;-W^2), ImZ_3^{(1)}(2;-W^2) $ are the imaginary parts 
of the first term of the expansion in $ (d-2) $ of the corresponding 
functions, see Eq.s(\ref{expd-2}) (let us recall once more that 
according to Eq.s(\ref{defZ22},\ref{defZ32}) $ Z_2(2;p^2), Z_3(2;p^2) $ 
vanish identically). An equation similar to Eq.(\ref{Im4SIm2a}) holds 
for $ ImS_1(4;-W^2) $; we do not write it explicitly for the sake of 
brevity. \\ 
The functions $ ImS(4;-W^2), ImS_1(4;-W^2) $ and 
$ ImS(2;-W^2), ImS_1(2;-W^2) $ are known, see Eq.s(\ref{Im4SI},\ref{Im4S1I}) 
and Eq.s(\ref{Im2SK},\ref{Im2S1K}); by combining Eq.(\ref{Im4SIm2a}) and 
the similar (not written) equation for $ ImS_1(4;-W^2) $, 
one can obtain the explicit values of 
$ ImZ_2^{(1)}(2;-W^2), ImZ_3^{(1)}(2;-W^2) $. One finds 
\begin{align}
 \frac{1}{\pi}ImZ_2^{(1)}(2;-W^2)  =  \frac{N_2}{16} &\Bigl[\,
        (W^2 - m_3^2 + m_2^2 - m_1^2)I(0,W)
       + I(1,W) \nonumber \\ 
       &- (m_3^2 - m_2^2)(W^2 - m_1^2)I(-1,W) \,\Bigr]\ , \labbel{Z12} 
\end{align} 
\begin{align}
 \frac{1}{\pi}ImZ_3^{(1)}(2;-W^2)  = \frac{N_2}{16} &\Bigl[\,
        (W^2 + m_3^2 - m_2^2 - m_1^2)I(0,W)
       +  I(1,W) \nonumber \\ 
       &+ (m_3^2 - m_2^2)(W^2 - m_1^2)I(-1,W) \, \Bigr]\ , \labbel{Z13} 
\end{align} 
From the previous equations and the same procedure giving 
Eq.s(\ref{Im2Sthr},\ref{Im2Sithr}) we obtain in particular the values at 
threshold 
\begin{align}
   \frac{1}{\pi}ImZ_2^{(1)}(2;-W^2) \xrightarrow[W \to (m_1+m_2+m_3) ]{} 
   & \frac{N_2}{16} \, \sqrt{\frac{m_2(m_1+m_2+m_3)}{m_1\,m_3}} \nonumber\\
   \frac{1}{\pi}ImZ_3^{(1)}(2;-W^2) \xrightarrow[W \to (m_1+m_2+m_3) ]{} 
   & \frac{N_2}{16} \, \sqrt{\frac{m_3(m_1+m_2+m_3)}{m_1\,m_2}} \ . 
\labbel{Z1k.thr}
\end{align} \par 
$ ImZ_2^{(1)}(2;-W^2) $ can also be evaluated solving, by quadrature, 
the imaginary part of the differential equation  Eq.(\ref{eqZ21d2}), 
{\it i.e.} by evaluating  
$$ Im Z_2^{(1)}(2;-W^2) = C +\ \int^{-W^2} dp^2 
      \ Im\Biggl( \frac{d}{dp^2} Z_2^{(1)}(2;p^2) \Biggr) \ , $$ 
where $ C $ is an integration constant and $ dZ_2^{(1)}(2;p^2)/dp^2 $ 
is obtained from Eq.(\ref{eqZ21d2}) itself. The constant $ C $ can be 
fixed, {\it a posteriori}, by requiring that the imaginary parts 
of the ``conventional" M.I. vanish at threshold in $ d=4 $ dimensions, 
a condition which leads again to Eq.s(\ref{Z1k.thr}). \\ 
After many algebraic simplifications, one obtains for $ ImZ_2^{(1)}(2;-W^2) $ 
\begin{align} 
 \frac{1}{\pi}ImZ_2^{(1)}(2;-W^2)  &= 
   \frac{N_2}{16} \sqrt{\frac{m_2(m_1+m_2+m_3) }{m_1 m_3} } \nonumber\\ 
 &+ \frac{1}{64}\int_{(m_1+m_2+m_3)^2}^{W^2}\, ds 
   \ \Bigl[ \ \ \tilde{F}(s,m_1,m_2,m_3)\ I(0,s) \nonumber\\ 
   & \hspace{3.8cm} -  \tilde{G}(s,m_1,m_2,m_3)\ I(1,s) \nonumber\\ 
   & \hspace{3.8cm} +  \tilde{H}(s,m_1,m_2,m_3)\ I(2,s) \Bigr] 
                                                    \ . \labbel{Z1k.a} 
\end{align} 
where the three quantities $ \tilde{F}, \tilde{G}, \tilde{H} $ 
are all expressed in terms of the corresponding functions $ F, G, H $ by 
the relation 
\begin{align} 
  \tilde{F}(s,m_1,m_2,m_3) &= 
                 F(s,m_1,m_2,m_3) + F(s,m_1,-m_2,m_3) \nonumber\\ 
              &+ F(s,m_1,m_2,-m_3) + F(s,m_1,-m_2,-m_3) \ , \nonumber 
\end{align} 
and the explicit expressions of those functions are
\begin{align} 
 F(s,m_1,m_2,m_3) &= \frac{(m_2+m_3)^2}{m_1m_3} 
         \ \ \frac{2m_1^2+m_2^2+m_3^2+2m_1m_2+2m_1m_3} 
                    {s-(m_1+m_2+m_3)^2} \ , \nonumber\\ 
 G(s,m_1,m_2,m_3) &= 2\frac{m_1^2+m_2^2+m_3^2+m_1m_2+m_1m_3+m_2m_3} 
                          {m_1m_3\ [s-(m_1+m_2+m_3)^2]} \ , \nonumber\\ 
 H(s,m_1,m_2,m_3) &= \frac{1}{m_1m_3\ [s-(m_1+m_2+m_3)^2]} \ . \nonumber 
\end{align} 
To carry out the integration, we use the integral representations 
Eq.(\ref{Inids}) for the elliptic integrals $ I(n,s) $ and exchange the 
order of integration according to 
$$ \int_{(m_1+m_2+m_3)^2}^{W^2} ds 
   \int_{(m_2+m_3)^2}^{(\sqrt{s}-m_1)^2} \frac{db}{\sqrtR} = 
   \int_{(m_2+m_3)^2}^{(W-m_1)^2} \frac{db}{\sqrt{R_2(b,m_2^2,m_3^2)}} 
   \int_{(\sqrt{b}+m_1)^2}^{W^2} \frac{ds}{\sqrt{R_2(s,b,m_1^2)}} \ , $$ 
where Eq.(\ref{R4R2R2}) was used. 
The $s $ integration is then elementary, giving only logarithms of 
suitable arguments and new square roots quadratic in $ b $\,; a subsequent 
integration by parts in $ b $ removes those logarithms with some of the 
accompanying square roots, and the result is Eq.(\ref{Z12}), as 
expected. \\ 
The same applies also for $ ImZ^{(1)}_3(2;-W^2) $, whose value is 
obtained by simply exchanging 
$ m_2 $ and $ m_3 $ in Eq.(\ref{Z12}). \\ 

\section{Conclusions}
\labbel{conc}

In this paper we introduced a new class of identities, dubbed Schouten 
identities, valid at fixed integer value of the dimensions $d$. 
We applied the identities valid at $ d=2 $ to the case of the massive 
two-loop sunrise graph with different masses, finding that 
in $d=2$ dimensions only two of the four Master Integrals (M.I.s) are 
actually independent, so that the other two can be expressed as
suitable linear combinations of the latter.

In the general case of arbitrary dimension $d$ and different masses, 
the four M.I.s are known to fulfil a system of
four first-order coupled differential equations in the external momentum
transfer. The system can equivalently be re-phrased as a fourth-order
differential equation for one of the M.I.s only.

Using these relations we introduced a new set of four 
independent M.I.s, valid for any number of dimensions $d$, 
whose property is that two of the newly defined integrals
vanish identically in $d=2$.
The new system of differential equations for this set of M.I.s 
takes then a simpler block form when expanded in $(d-2)$.

Starting from this system, one can derive a second-order differential
equation, \textsl{exact in $d$}, for the full scalar amplitude, which
still contains the two integrals, whose value is zero at $d=2$,
as inhomogeneous terms.
We verified that the zeroth-order of our equation corresponds to
the equation derived in~\cite{Weinzierl2012}.
Our equations, once expanded in powers of $(d-2)$, can be used,
together with the linear equations for the remaining three M.I.s,
for evaluating recursively, at least in principle, all four M.I.s, 
up to any order in $(d-2)$.

We then worked out explicitly the Tarasov's shifting relations needed to 
recover the physically more relevant value of the four M.I.s expanded 
in $(d-4)$ at $ d\approx4$ starting from the expansion in $(d-2)$ at 
$d\approx2$ worked out in our approach. 

As an example of this procedure we discussed the relationship
between the \textsl{imaginary parts} of the four M.I.s in
$d=2$ and $d=4$.
The latter can be computed using the Cutkosky-Veltman rule.
We showed how in $d=2$ the imaginary parts of the four M.I.s can
be written in terms of \textsl{two} independent functions only,
namely the complete elliptic integrals of the first and of the second kind.
The same is not true in $d=4$ dimensions, where four independent
elliptic integrals are needed in order to represent the four imaginary parts.
We then showed how the Tarasov's shift formulas 
relate the imaginary parts in $d=2$ and $d=4$ dimensions.
Finally, we gave an explicit example of how
the differential equations for the imaginary
parts of the master integrals can be integrated by quadrature.

\section*{Acknowledgements} 
We are grateful to J.\,Vermaseren for his assistance in the use of 
the algebraic program FORM~\cite{form} which was intensively used in 
all the steps of the calculation, to T.\,Gehrmann for many interesting comments
and discussions, and to F.\,Cascioli for proofreading the whole manuscript.
L.T. wishes to thank
A.\,von Manteuffel for his assistance with Reduze 2~\cite{Reduze2},
with which the reduction to Master Integrals of 
the two-loop sunrise has been carried out, and
G.\,Heinrich and S.\,Borowka for their help with SecDec~\cite{SecDec2}.
This research was supported in part by the 
Swiss National Science Foundation (SNF) under the contract PDFMP2-135101.

\appendix 

\section{The polynomials of the first-order differential 
equations}
\labbel{App:Poly1}

In this appendix we give the explicit expressions for the polynomials
appearing in the first-order differential equations in section~\ref{NSMI}.
All polynomials are functions of $p^2$ and of the three masses $m_1$,
$m_2$, $m_3$, while they do not depend on the dimensions $d$.

\begin{align}
 P_{10}^{(0)}(p^2,m_1,m_2,m_3) &= 
 - m_1^2 (m_1+m_2+m_3) (m_1-m_2+m_3) (m_1+m_2-m_3) (m_1-m_2-m_3)
 \nonumber \\ &\times 
        \left( m_1^2-m_2^2-m_3^2 \right)^2  \nonumber \\
       &+ p^2    \left(  m_1^8 + 4 m_2^2 m_1^6 - 14 m_2^4 m_1^4 
       + 12 m_2^6 m_1^2 - 3 m_2^8 + 4 
         m_3^2 m_1^6 + 4 m_3^2 m_2^4 m_1^2 
         \right. \nonumber \\ & \left.- 8 m_3^2 m_2^6 
         - 14 m_3^4 m_1^4 + 4 m_3^4 
         m_2^2 m_1^2 + 22 m_3^4 m_2^4 + 12 m_3^6 m_1^2 
         - 8 m_3^6 m_2^2 - 3 m_3^8  \right)\nonumber \\
       &+ p^4    \left(  10 m_1^6 - 4 m_2^2 m_1^4 + 2 m_2^4 m_1^2 
       - 8 m_2^6 - 4 m_3^2 m_1^4
          + 16 m_3^2 m_2^4 + 2 m_3^4 m_1^2 
          \right. \nonumber \\ & \left.+ 16 m_3^4 m_2^2 - 8 m_3^6  \right)
          \nonumber \\
       &+ p^6    \left(  14 m_1^4 - 4 m_2^2 m_1^2 - 6 m_2^4 
       - 4 m_3^2 m_1^2 + 24 m_3^2 m_2^2
          - 6 m_3^4  \right)\nonumber \\
       &+ 7 p^8 \, m_1^2 
       + p^{10}\,,
\end{align}

\begin{align}
 P_{10}^{(1)}(p^2,m_1,m_2,m_3) &=  (m_1+m_2+m_3) (m_1-m_2+m_3) 
 (m_1+m_2-m_3) (m_1-m_2-m_3)
 \nonumber \\ &\times  \left( m_1^2-m_2^2-m_3^2 \right)    
 \left(  7 m_1^4 - 6 m_2^2 m_1^2 - m_2^4 - 6 m_3^2 m_1^2 
 + 2 m_3^2 m_2^2 - m_3^4  \right)\nonumber \\
       &+ p^2    \left(  11 m_1^8 - 48 m_2^2 m_1^6 
       + 78 m_2^4 m_1^4 - 56 m_2^6 m_1^2 + 15 m_2^8
          - 48 m_3^2 m_1^6 \right. \nonumber \\ & \left.
          + 68 m_3^2 m_2^2 m_1^4 - 40 m_3^2 m_2^4 m_1^2 
          + 20 m_3^2 m_2^6 + 78 m_3^4 m_1^4 
          - 40 m_3^4 m_2^2 m_1^2 \right. \nonumber \\ & \left.
          - 70 m_3^4 m_2^4 - 56 m_3^6 m_1^2
          + 20 m_3^6 m_2^2 + 15 m_3^8  \right)\nonumber \\
       &+ p^4    \left(   - 2 m_1^6 - 14 m_2^2 m_1^4 + 2 m_2^4 m_1^2 
       + 14 m_2^6 - 14 m_3^2 
         m_1^4 + 60 m_3^2 m_2^2 m_1^2 
         \right. \nonumber \\ & \left.- 62 m_3^2 m_2^4 + 2 m_3^4 m_1^2 
         - 62 m_3^4 m_2^2
          + 14 m_3^6  \right)\nonumber \\
       &- 2\, p^6  \left( m_1^2-m_2^2-3 m_3^2 \right)  
       \left( m_1^2-3 m_2^2-m_3^2 \right)  
       + p^8    \left(  11 m_1^2 + m_2^2 + m_3^2  \right)
       + 7\,p^{10}\,,
\end{align}

\begin{align}
 P_{10}^{(2)}(p^2,m_1,m_2,m_3) &= (m_1+m_2+m_3) (m_1-m_2+m_3) 
 (m_1+m_2-m_3) (m_1-m_2-m_3)\nonumber \\ &\times   
         \left(  5 m_1^4 - 4 m_2^2 m_1^2 - m_2^4 
         - 4 m_3^2 m_1^2 + 2 m_3^2 m_2^2 - m_3^4  \right) \nonumber \\
       &+ p^2    \left(  8 m_1^6 - 18 m_2^2 m_1^4 
       + 20 m_2^4 m_1^2 - 10 m_2^6 - 18 m_3^2 m_1^4
          + 24 m_3^2 m_2^2 m_1^2 \right. \nonumber \\ & \left.
          + 10 m_3^2 m_2^4 + 20 m_3^4 m_1^2 + 10 m_3^4 m_2^2 - 
         10 m_3^6  \right) \nonumber \\
       &+ p^4    \left(  10 m_1^4 + 6 m_2^2 m_1^2 - 8 m_2^4 
       + 6 m_3^2 m_1^2 + 48 m_3^2 m_2^2
          - 8 m_3^4  \right) \nonumber \\
       &+ p^6    \left(  16 m_1^2 + 10 m_2^2 + 10 m_3^2  \right)
       +9\, p^8\,,  
\end{align}

\begin{align}
    P_{11}^{(0)}(p^2,m_1,m_2,m_3) &=
       (m_1+m_2+m_3)^2 (m_1-m_2+m_3)^2 
       (m_1+m_2-m_3)^2 (m_1-m_2-m_3)^2\nonumber \\ &\times 
          \left( m_1^2-m_2^2-m_3^2 \right) \, m_1^2\nonumber \\
       &+ p^2    \left(   - 6 m_1^{10} + m_2^2 m_1^8 + 32 m_2^4 m_1^6 
       - 42 m_2^6 m_1^4 + 14 m_2^8
          m_1^2 + m_2^{10} + m_3^2 m_1^8 
          \right. \nonumber \\ & \left.- 64 m_3^2 m_2^2 m_1^6 
          + 26 m_3^2 m_2^4 m_1^4 + 
         40 m_3^2 m_2^6 m_1^2 - 3 m_3^2 m_2^8 + 32 m_3^4 m_1^6
          \right. \nonumber \\ & \left.+ 26 m_3^4 m_2^2 m_1^4 - 
         108 m_3^4 m_2^4 m_1^2 + 2 m_3^4 m_2^6 - 42 m_3^6 m_1^4 
         + 40 m_3^6 m_2^2 m_1^2
          \right. \nonumber \\ & \left.+ 2 m_3^6 m_2^4 + 14 m_3^8 m_1^2 
          - 3 m_3^8 m_2^2 + m_3^{10}  \right)\nonumber \\
       &+ p^4    \left(   - 33 m_1^8 + 6 m_2^2 m_1^6 - 20 m_2^4 m_1^4 
       + 42 m_2^6 m_1^2 + 5 
         m_2^8 + 6 m_3^2 m_1^6 \right. \nonumber \\ & \left.- 24 m_3^2 m_2^2 m_1^4 
         - 26 m_3^2 m_2^4 m_1^2 - 4 m_3^2 
         m_2^6 - 20 m_3^4 m_1^4 - 26 m_3^4 m_2^2 m_1^2 
         \right. \nonumber \\ & \left.- 2 m_3^4 m_2^4 + 42 m_3^6 m_1^2
          - 4 m_3^6 m_2^2 + 5 m_3^8  \right)\nonumber \\
       &+ p^6    \left(   - 52 m_1^6 - 6 m_2^2 m_1^4 + 32 m_2^4 m_1^2 
       + 10 m_2^6 - 6 m_3^2 
         m_1^4 - 64 m_3^2 m_2^2 m_1^2 \right. \nonumber \\ & \left.
         + 6 m_3^2 m_2^4 + 32 m_3^4 m_1^2 + 6 m_3^4 m_2^2
          + 10 m_3^6  \right)\nonumber \\
       &+ p^8    \left(   - 33 m_1^4 - m_2^2 m_1^2 + 10 m_2^4 
       - m_3^2 m_1^2 + 12 m_3^2 m_2^2
          + 10 m_3^4  \right)\nonumber \\
       &+ p^{10}    \left(   - 6 m_1^2 + 5 m_2^2 + 5 m_3^2  \right)
       + p^{12}\,,  
\end{align}

\begin{align}
    P_{11}^{(1)}(p^2,m_1,m_2,m_3) &=
        (m_1+m_2+m_3)^2 (m_1-m_2+m_3)^2 (m_1+m_2-m_3)^2
         (m_1-m_2-m_3)^2 \nonumber \\ &\times   \left(  5 m_1^4 - 4 
         m_2^2 m_1^2 - m_2^4 - 4 m_3^2 m_1^2 
         + 2 m_3^2 m_2^2 - m_3^4  \right)\nonumber \\
       &+ p^2    \left(  2 m_1^{10} - 28 m_2^2 m_1^8 + 76 m_2^4 m_1^6 
       - 80 m_2^6 m_1^4 + 34 m_2^8
          m_1^2 \right. \nonumber \\ & \left.- 4 m_2^{10} - 28 m_3^2 m_1^8 
          + 40 m_3^2 m_2^2 m_1^6 - 48 m_3^2 m_2^4 
         m_1^4 + 24 m_3^2 m_2^6 m_1^2 \right. \nonumber \\ & \left.
         + 12 m_3^2 m_2^8 + 76 m_3^4 m_1^6 - 48 m_3^4 m_2^2
          m_1^4 - 116 m_3^4 m_2^4 m_1^2 - 8 m_3^4 m_2^6 
          \right. \nonumber \\ & \left.- 80 m_3^6 m_1^4 + 24 m_3^6 
         m_2^2 m_1^2 - 8 m_3^6 m_2^4 + 34 m_3^8 m_1^2 
         + 12 m_3^8 m_2^2 - 4 m_3^{10}  \right)\nonumber \\
       &+ p^4    \left(   - 41 m_1^8 - 42 m_2^4 m_1^4 
       + 88 m_2^6 m_1^2 - 5 m_2^8 + 52 m_3^2 
         m_2^2 m_1^4 - 152 m_3^2 m_2^4 m_1^2 
         \right. \nonumber \\ & \left.+ 4 m_3^2 m_2^6 - 42 m_3^4 m_1^4 
         - 152 m_3^4 m_2^2 m_1^2 + 2 m_3^4 m_2^4 + 88 m_3^6 m_1^2 
         + 4 m_3^6 m_2^2 - 5 m_3^8  \right)\nonumber \\
       &+ p^6    \left(   - 84 m_1^6 - 8 m_2^2 m_1^4 + 60 m_2^4 m_1^2 
       - 8 m_3^2 m_1^4 - 184 m_3^2 m_2^2 m_1^2 + 60 m_3^4 m_1^2 
        \right)\nonumber \\
       &+ p^8    \left(   - 61 m_1^4 - 8 m_2^2 m_1^2 + 5 m_2^4 
       - 8 m_3^2 m_1^2 + 6 m_3^2 
         m_2^2 + 5 m_3^4  \right)\nonumber \\
       &+ p^{10}    \left(   - 14 m_1^2 + 4 m_2^2 + 4 m_3^2  \right)
       + p^{12}\,,  
\end{align}

\begin{align}
    P_{12}^{(0)}(p^2,m_1,m_2,m_3) &=
        (m_1+m_2+m_3) (m_1-m_2+m_3) (m_1+m_2-m_3) 
        (m_1-m_2-m_3)\nonumber \\ &\times  
        \left( m_1^2-m_2^2-m_3^2 \right)    \left(  3 
         m_1^2 + m_2^2 - m_3^2  \right)\nonumber \\
       &+ p^2    \left(  18 m_1^6 + 2 m_2^2 m_1^4 - 10 m_2^4 m_1^2 
       - 10 m_2^6 - 32 m_3^2 m_1^4
          + 24 m_3^2 m_2^2 m_1^2 \right. \nonumber \\& \left.
          + 40 m_3^2 m_2^4 + 34 m_3^4 m_1^2 - 10 m_3^4 m_2^2 - 
         20 m_3^6  \right)\nonumber \\
       &+ p^4    \left(  36 m_1^4 - 12 m_2^2 m_1^2 - 8 m_2^4 
       + 6 m_3^2 m_1^2 + 66 m_3^2 m_2^2
          - 34 m_3^4  \right)\nonumber \\
       &+ p^6    \left(  30 m_1^2 + 10 m_2^2 - 4 m_3^2  \right)
       + 9\,p^8\,,
\end{align}

\begin{align}
   P_{14}^{(1)}(p^2,m_1,m_2,m_3) &=
        (m_1+m_2+m_3) (m_1-m_2+m_3) (m_1+m_2-m_3) 
        (m_1-m_2-m_3) \nonumber \\ &\times  \left( m_1^2-m_2^2-m_3^2 \right) 
       \left( m_1^2-m_2^2+m_3^2 \right) \nonumber \\
       &+ p^2    \left(  6 m_1^6 - 22 m_2^2 m_1^4 + 26 m_2^4 m_1^2 
       - 10 m_2^6 + 12 m_3^2 m_1^4
          + 8 m_3^2 m_2^2 m_1^2 - 20 m_3^2 m_2^4 
          \right. \nonumber \\  &\left.- 18 m_3^4 m_1^2 + 30 m_3^4 m_2^2  \right)
          \nonumber \\
       &+ p^4    \left(  12 m_1^4 + 8 m_2^2 m_1^2 - 20 m_2^4
        - 10 m_3^2 m_1^2 + 22 m_3^2 
         m_2^2 + 6 m_3^4  \right)\nonumber \\
       &+ p^6    \left(  10 m_1^2 - 6 m_2^2 + 8 m_3^2  \right)
       + 3\,p^8 \,,
\end{align}

\begin{align}
    P_{14}^{(2)}(p^2,m_1,m_2,m_3) &=
        (m_1+m_2+m_3) (m_1-m_2+m_3) (m_1+m_2-m_3)
         (m_1-m_2-m_3) \nonumber \\ &\times  \left( m_1^2-m_2^2+m_3^2 \right)   
          \left(  5  m_1^4 - 4 m_2^2 m_1^2 - m_2^4 - 4 m_3^2 m_1^2 
          + 2 m_3^2 m_2^2 - m_3^4  \right) \nonumber \\
      &+ p^2  \left( m_1^2-m_2^2+m_3^2 \right)    
      \left(  17 m_1^6 - 31 m_2^2 m_1^4 + 19 m_2^4 m_1^2 - 5 
         m_2^6 - 29 m_3^2 m_1^4 
         \right. \nonumber \\ & \left.+ 46 m_3^2 m_2^2 m_1^2 
         + 7 m_3^2 m_2^4 + 15 m_3^4 m_1^2
          + m_3^4 m_2^2 - 3 m_3^6  \right) \nonumber \\
      &+ p^4    \left(  22 m_1^6 + 14 m_2^2 m_1^4 - 46 m_2^4 m_1^2 
      + 10 m_2^6 - 42 m_3^2 m_1^4 + 72 m_3^2 m_2^2 m_1^2
          \right. \nonumber \\ & \left.- 6 m_3^2 m_2^4 + 38 m_3^4 m_1^2 
          - 2 m_3^4 m_2^2 - 2 m_3^6  \right) \nonumber \\
      &+ p^6    \left(  14 m_1^4 - 16 m_2^2 m_1^2 
      + 10 m_2^4 + 28 m_3^2 m_1^2 + 4 m_3^2 
         m_2^2 + 2 m_3^4  \right) \nonumber \\
      &+ p^8    \left(  5 m_1^2 + 5 m_2^2 + 3 m_3^2  \right)
       + p^{10} \,,
\end{align}

\begin{align}
    P_{22}(p^2,m_1,m_2,m_3) &=  3 m_1^4 - 2 m_2^2 m_1^2 
    - m_2^4 - 2 m_3^2 m_1^2 + 2 m_3^2 m_2^2 - m_3^4 \nonumber \\
       &+ 2\,p^2  \left( m_1^2-m_2^2+m_3^2 \right)
       + 3\,p^4\,, 
\end{align}

The polynomials defined above fulfil, among the others, the relation:
\begin{align}\label{eq:polynomials}
  P_{14}^{(2)}(p^2,m_1,m_2,m_3) + P_{14}^{(2)}(p^2,m_1,m_3,m_2) 
       - 2 m_1^2 P_{10}^{(2)}(p^2,m_1,m_2,m_3)
       - 2 p^2 P^2(p^2,m_1,m_2,m_3) = 0\,,
\end{align}
where note that the polynomial $P(p^2,m_1,m_2,m_3)$,
defined in Eq.(\ref{defP}),  appears squared.

\section{The polynomials of the second-order differential equation.}
\labbel{App:Poly2} 
In this second appendix we give the explicit expressions of the polynomials 
that appear in the second-order differential equation derived in section \ref{IIordDE}.
Also in this case, they are functions of $p^2$ and of the three masses $m_1$,
$m_2$ and $m_3$, but they do not depend on the dimensions $d$.

\begin{align} 
A_2^{(0)}(p^2,m_1,m_2,m_3) =  &- 
 (m_1 - m_2 - m_3)^3 (m_1 - m_2 + m_3)^3  (m_1 + m_2 - m_3)^3 
  (m_1 +  m_2 + m_3)^3\nonumber \\
       & - 8  \,p^{2}   
          (m_1 - m_2 - m_3)  (m_1 - m_2 + m_3)  (m_1 + m_2 - m_3) 
           (m_1 + m_2 + m_3)\nonumber \\ &\times  
           \left(m_1^6 - m_2^2 m_1^4 - m_2^4 m_1^2 + m_2^6 - m_3^2 m_1^4 + 10 
         m_3^2 m_2^2 m_1^2 \right. \nonumber \\ 
         &\left.- m_3^2 m_2^4 - m_3^4 m_1^2 - m_3^4 m_2^2 + m_3^6 \right)
          \nonumber \\
       &- p^{4}   
           \left(13 m_1^8 - 36 m_2^2 m_1^6 + 46 m_2^4 m_1^4 - 36 m_2^6 m_1^2 
           + 13 m_2^8 - 36 m_3^2 m_1^6 \right. \nonumber \\ 
          &- 124 m_3^2 m_2^2 m_1^4 - 124 m_3^2 m_2^4 m_1^2 - 36 m_3^2 
         m_2^6 + 46 m_3^4 m_1^4 - 124 m_3^4 m_2^2 m_1^2 
         \nonumber \\ &\left.+ 46 m_3^4 m_2^4 - 36 m_3^6 
         m_1^2 - 36 m_3^6 m_2^2 + 13 m_3^8\right)         
          \nonumber \\ &+8 \,p^{6}   
           \left(m_1^2 + m_2^2 + m_3^2\right) 
            \left(m_1^4 + 6 m_2^2 m_1^2 + m_2^4 + 6 m_3^2 
         m_1^2 + 6 m_3^2 m_2^2 + m_3^4\right)     \nonumber \\
       &+ p^{8}   
           \left(37 m_1^4 + 70 m_2^2 m_1^2 + 37 m_2^4 
           + 70 m_3^2 m_1^2 + 70 m_3^2 m_2^2
          + 37 m_3^4\right) \nonumber \\
       &+ 32  \,p^{10}   
           \left(m_1^2 + m_2^2 + m_3^2\right)
       + 9 \,p^{12}   
\end{align}

\begin{align} 
A_2^{(1)}(p^2,m_1,m_2,m_3) = &- \frac{1}{2}\,  
           (m_1 - m_2 - m_3)^3  (m_1 - m_2 + m_3)^3  (m_1 + m_2 - m_3)^3 
          (m_1 + m_2 + m_3)^3 \nonumber \\
       &+ p^{2}   
           (m_1 - m_2 - m_3)  (m_1 - m_2 + m_3)  (m_1 + m_2 - m_3)  (m_1 + m_2
          + m_3) \nonumber \\ &\times 
        \left(5 m_1^6 - 5 m_2^2 m_1^4 - 5 m_2^4 m_1^2 + 5 m_2^6 - 5 m_3^2 m_1^4
          + 2 m_3^2 m_2^2 m_1^2 \right. \nonumber \\ &\left.
          - 5 m_3^2 m_2^4 - 5 m_3^4 m_1^2 - 5 m_3^4 m_2^2 + 5 
         m_3^6\right) \nonumber \\
       &+ \frac{1}{2}\,  p^{4}  
            \left(41 m_1^8 - 84 m_2^2 m_1^6 + 86 m_2^4 m_1^4 - 84 m_2^6 m_1^2 + 41 
         m_2^8 - 84 m_3^2 m_1^6 \right.
          \nonumber \\ &+ 52 m_3^2 m_2^2 m_1^4 
          + 52 m_3^2 m_2^4 m_1^2 - 84 m_3^2
          m_2^6 + 86 m_3^4 m_1^4 + 52 m_3^4 m_2^2 m_1^2 
          \nonumber \\ &\left. + 86 m_3^4 m_2^4 - 84 m_3^6 
         m_1^2 - 84 m_3^6 m_2^2 + 41 m_3^8\right) \nonumber \\
       &+ 2  p^{6}   
           \left(11 m_1^6 - 19 m_2^2 m_1^4 
           - 19 m_2^4 m_1^2 + 11 m_2^6 - 19 m_3^2 
         m_1^4 + 54 m_3^2 m_2^2 m_1^2 
         \right. \nonumber \\ &\left. 
        - 19 m_3^2 m_2^4 - 19 m_3^4 m_1^2 - 19 m_3^4 m_2^2
          + 11 m_3^6\right)\nonumber \\
       &+ \frac{1}{2}\,  p^{8}   
            \left(m_1^4 - 50 m_2^2 m_1^2 + m_2^4 - 50 m_3^2 m_1^2 - 50 m_3^2 m_2^2
          + m_3^4\right) \nonumber \\
       &- 11 p^{10}   
            \left(m_1^2 + m_2^2 + m_3^2\right)
       - \frac{9}{2}\,  p^{12}
\end{align}

\begin{align} 
A_3^{(0)}(p^2,m_1,m_2,m_3) &= 
   (m_1 - m_2 - m_3) (m_1 - m_2 + m_3) (m_1 + m_2 - m_3) (m_1 + m_2 + m_3) 
\nonumber \\ 
&\times \left(m_1^6 - m_2^2 m_1^4 - m_2^4 m_1^2 + m_2^6 - m_3^2 m_1^4 \right.
       \nonumber \\ &\left.+ 6 m_3^2 m_2^2 
      m_1^2 - m_3^2 m_2^4 - m_3^4 m_1^2 - m_3^4 m_2^2 + m_3^6\right)
      \nonumber \\
       &+ p^{2}   
   \left(5 m_1^8 - 8 m_2^2 m_1^6 + 6 m_2^4 m_1^4 
   - 8 m_2^6 m_1^2 + 5 m_2^8
    - 8  m_3^2 m_1^6 \right.\nonumber \\ &
    - 8 m_3^2 m_2^2 m_1^4 - 8 m_3^2 m_2^4 m_1^2 - 8 m_3^2 m_2^6 + 6 
         m_3^4 m_1^4 - 8 m_3^4 m_2^2 m_1^2 
   \nonumber \\ &\left.+ 6 m_3^4 m_2^4 - 8 m_3^6 m_1^2 - 8 m_3^6 
         m_2^2 + 5 m_3^8\right)
          \nonumber \\
       &+ 2 p^{4}   
           \left(3 m_1^6 - 7 m_2^2 m_1^4 - 7 m_2^4 m_1^2 + 3 m_2^6 - 7 m_3^2 m_1^4 
                 \right.\nonumber \\ &\left.- 
         7 m_3^2 m_2^4 - 7 m_3^4 m_1^2 - 7 m_3^4 m_2^2 + 3 m_3^6\right)
          \nonumber \\
       & - 2 p^{6}   
           \left(m_1^4 + 8 m_2^2 m_1^2 + m_2^4 
           + 8 m_3^2 m_1^2 + 8 m_3^2 m_2^2 + m_3^4 \right)
         \nonumber \\   
       &- 7 p^{8}   
           \left(m_1^2 + m_2^2 + m_3^2\right)
       - 3 p^{10}  
\end{align}

\begin{align} 
A_3^{(1)}(p^2,m_1,m_2,m_3) =  &- \frac{1}{2}\, 
           (m_1 - m_2 - m_3) (m_1 - m_2 + m_3) (m_1 + m_2 - m_3) (m_1+ m_2 + m_3)
\nonumber \\ &\times \left(m_1^2 - m_2^2 - m_3^2\right) \left(m_1^2 - m_2^2 + m_3^2\right) 
                     \left(m_1^2 + m_2^2 - m_3^2\right)
          \nonumber \\
       &- \frac{1}{2}\,  p^{2}   
          \left(17 m_1^8 - 32 m_2^2 m_1^6 + 18 m_2^4 m_1^4 - 8 m_2^6 m_1^2 + 5 
         m_2^8 - 32 m_3^2 m_1^6 \right. \nonumber \\ 
         &+ 20 m_3^2 m_2^2 m_1^4 + 8 m_3^2 m_2^4 m_1^2 + 4 m_3^2 
         m_2^6 + 18 m_3^4 m_1^4 + 8 m_3^4 m_2^2 m_1^2 \nonumber \\ 
        &\left. - 18 m_3^4 m_2^4 - 8 m_3^6 m_1^2
          + 4 m_3^6 m_2^2 + 5 m_3^8\right)
          \nonumber \\
       &- p^{4}   
           \left(21 m_1^6 - 31 m_2^2 m_1^4 + 7 m_2^4 m_1^2 + 3 m_2^6 - 31 m_3^2 m_1^4
          + 30 m_3^2 m_2^2 m_1^2 \right. \nonumber \\ 
          &\left. - 3 m_3^2 m_2^4 + 7 m_3^4 m_1^2 - 3 m_3^4 m_2^2 + 3 
         m_3^6\right)
          \nonumber \\
       &- p^{6}   
           \left(17 m_1^4 - 20 m_2^2 m_1^2 - m_2^4 - 20 m_3^2 m_1^2 + 22 m_3^2 m_2^2 - 
         m_3^4\right)
          \nonumber \\
       & - \frac{1}{2}\, p^{8}   
           \left(5 m_1^2 - 7 m_2^2 - 7 m_3^2\right)
       + \frac{3}{2}\,   p^{10}  
\end{align}

\begin{align} 
A_4(p^2,m_1,m_2,m_3) &= - \left(m_1^2 - m_2^2\right) \, \Big[ \nonumber \\
           &24 (m_1 - m_2 - m_3) (m_1 - m_2 + m_3) (m_1 + m_2- m_3) (m_1 + m_2 + m_3)
             \nonumber \\ &\times   \left(m_1^2 + m_2^2 - m_3^2 \right)\nonumber \\
       &+ 8 p^{2}   
            \left(9 m_1^4 - 10 m_2^2 m_1^2 + 9 m_2^4 - 14 m_3^2 
         m_1^2 - 14 m_3^2 m_2^2 + 5 m_3^4\right)
          \nonumber \\
       &+ 24  p^{4}   
           \left(3 m_1^2 + 3 m_2^2 - 7 m_3^2\right) 
       + 24 p^{6} \,\Big]
\end{align}

\begin{align} 
A_5^{(1)}(p^2,m_1,m_2,m_3) &=  
           (m_1 - m_2 - m_3) (m_1 - m_2 + m_3) (m_1 + m_2 - m_3) (m_1 + m_2
          + m_3)\nonumber \\ &\times \left(m_1^4 - 2 m_2^2 m_1^2 + m_2^4 + m_3^2 m_1^2 
                                     + m_3^2 m_2^2 - 2 m_3^4\right)\nonumber \\
       &+ p^{2}   
           \left(3 m_1^6 - 3 m_2^2 m_1^4 - 3 m_2^4 m_1^2 + 3 m_2^6 - 8 m_3^2 m_1^4 - 8 
         m_3^2 m_2^4 \right.\nonumber \\
         &\left.+ 11 m_3^4 m_1^2 + 11 m_3^4 m_2^2 - 6 m_3^6\right)\nonumber \\
       &+ p^{4}   
          \left(3 m_1^4 - 14 m_2^2 m_1^2 + 3 m_2^4 + 7 m_3^2 m_1^2 + 7 m_3^2 m_2^2 - 6
          m_3^4\right)\nonumber \\
       &+ p^{6}   
          \left(m_1^2 + m_2^2 - 2 m_3^2\right)
\end{align}

\begin{align} 
A_5^{(2)}(p^2,m_1,m_2,m_3) =
&- \frac{1}{2}\, (m_1 - m_2 - m_3) (m_1 - m_2 + m_3) (m_1 + m_2 - m_3) (m_1+ m_2 + m_3)
              \nonumber \\ &\times \left(m_1^2 - m_2^2 - m_3^2\right) 
                                   \left(m_1^2 - m_2^2 + m_3^2\right)\nonumber \\
       &- p^{2} 
           \left(3 m_1^6 - 3 m_2^2 m_1^4 - 3 m_2^4 m_1^2 + 3 m_2^6 - 2 m_3^2 m_1^4 + 4 
         m_3^2 m_2^2 m_1^2 \right. \nonumber \\ &\left. 
        - 2 m_3^2 m_2^4 + 7 m_3^4 m_1^2 + 7 m_3^4 m_2^2 - 8 m_3^6\right)
          \nonumber \\
       &- p^{4}   
           \left(6 m_1^4 - 12 m_2^2 m_1^2 + 6 m_2^4 + 11 m_3^2 m_1^2 + 11 m_3^2 m_2^2
          - 13 m_3^4\right)
          \nonumber \\
       &- p^{6}   
           \left(5 m_1^2 + 5 m_2^2 - 4 m_3^2\right)
       - \frac{3}{2}\,   p^{8}   
\end{align}

Note that, in order to derive Eq.(\ref{eq:IIord2}), 
we made use of the following relation:
\be \labbel{relApol}
A_5^{(1)}(p^2,m_1,m_2,m_3) + A_5^{(1)}(p^2,m_1,m_3,m_2) 
+ A_5^{(1)}(p^2,m_2,m_3,m_1) = 0\,. 
\ee

\section{Tarasov's shift}
\labbel{App:Tar}

In this Appendix we enclose the explicit formula for the order zero of the Tarasov's shift,
Eq.(\ref{eq:tarinv}) discussed in section \ref{Shift}, which relates the zeroth-order
of the full scalar amplitude, evaluated in $d=4$ dimensions, to a linear combination of 
the four new M.I.s evaluate in $d=2$ dimensions, namely
$S(2;p^2)$, $S_1(2;p^2)$, $Z_2^{(1)}(2;p^2)$ and $Z_3^{(1)}(2;p^2)$.

\begin{align}
   S^{(0)}(4;p^2) &= - \frac{1}{128} \left[ 13 p^2 + 24 (m_1^2 + m_2^2 + m_3^2) \right]\nonumber \\
          &+ \frac{1}{8} \Big[ (m_1^2 - m_2^2 - m_3^2) \ln{(m_3)} \ln{(m_2)}
           -  (m_1^2 - m_2^2 + m_3^2) \ln{(m_3)} \ln{(m_1)}\nonumber \\
          &-  (m_1^2 + m_2^2 - m_3^2) \ln{(m_2)} \ln{(m_1)}
          -  \ln{(m_3)}^2 m_3^2
          -  \ln{(m_2)}^2 m_2^2
          -  \ln{(m_1)}^2 m_1^2 \Big] \nonumber \\
       &+ \frac{1}{96\,p^2} \Big\{
           \Big[ 2 p^4 + 6 (4 m_1^2 + m_2^2 + m_3^2) p^2 \nonumber \\&+ (2 m_1^4 - 6 
         m_2^2 m_1^2 - m_2^4 - 6 m_3^2 m_1^2 + 12 m_3^2 m_2^2 - m_3^4)\Big] \ln{(m_1)}\nonumber \\
          &+  \Big[ 2 p^4 + 6 (m_1^2 + 4 m_2^2 + m_3^2) p^2 \nonumber \\&- (m_1^4 + 6
        m_2^2 m_1^2 - 2 m_2^4 - 12 m_3^2 m_1^2 + 6 m_3^2 m_2^2 + m_3^4)\Big] \ln{(m_2)} \nonumber \\
          &+  \Big[ 2 p^4 + 6 (m_1^2 + m_2^2 + 4 m_3^2) p^2 \nonumber \\&- (m_1^4 - 
       12 m_2^2 m_1^2 + m_2^4 + 6 m_3^2 m_1^2 + 6 m_3^2 m_2^2 - 2 m_3^4)\Big] \ln{(m_3)} \Big]
        \Big\} \nonumber 
        \\
       &- \frac{1}{96\,p^2\,P(p^2,m_1,m_2,m_3)}  \Big\{
            \Big[ 2 p^8 - 2 (2 m_1^2 - 5 m_2^2 - 5 m_3^2) p^6 \nonumber \\&+ (26 
         m_1^4 - 56 m_2^2 m_1^2 + 13 m_2^4 - 56 m_3^2 m_1^2 + 32 m_3^2 m_2^2 + 13 m_3^4
         ) p^4 \nonumber \\&+ 2 (16 m_1^6 - 25 m_2^2 m_1^4 - 17 m_2^4 m_1^2 + 2 m_2^6 - 25 
         m_3^2 m_1^4 + 8 m_3^2 m_2^4 \nonumber \\&- 17 m_3^4 m_1^2 + 8 m_3^4 m_2^2 + 2 m_3^6) p^2 
       \nonumber \\&+ 
         (16 m_2^2 m_1^6 - 13 m_2^4 m_1^4 - 2 m_2^6 m_1^2 - m_2^8 + 16 m_3^2 m_1^6
          - 100 m_3^2 m_2^2 m_1^4 + 22 m_3^2 m_2^4 m_1^2 \nonumber \\&+ 14 m_3^2 m_2^6 - 13 m_3^4 
         m_1^4 + 22 m_3^4 m_2^2 m_1^2 - 26 m_3^4 m_2^4 - 2 m_3^6 m_1^2 + 14 m_3^6 m_2^2
          - m_3^8) \Big] \ln{(m_1)}\nonumber   \displaybreak \\
          &-  \Big[p^8 - 2 (m_1^2 - 7 m_2^2 + 2 m_3^2) p^6 \nonumber \\&+ (13 m_1^4
          - 22 m_2^2 m_1^2 + 26 m_2^4 - 34 m_3^2 m_1^2 + 16 m_3^2 m_2^2 - 13 m_3^4) 
         p^4 \nonumber \\&+ 2 (8 m_1^6 - 50 m_2^2 m_1^4 + 11 m_2^4 m_1^2 + 7 m_2^6 + 25 m_3^2 
         m_1^4 - 8 m_3^2 m_2^4 \nonumber \\&- 28 m_3^4 m_1^2 + 16 m_3^4 m_2^2 - 5 m_3^6) p^2 
         \nonumber \\&+ (
          - 16 m_2^2 m_1^6 + 13 m_2^4 m_1^4 + 2 m_2^6 m_1^2 + m_2^8 + 32 m_3^2 m_1^6 - 
         50 m_3^2 m_2^2 m_1^4 - 34 m_3^2 m_2^4 m_1^2 
         \nonumber \\&+ 4 m_3^2 m_2^6 - 26 m_3^4 m_1^4 + 
         56 m_3^4 m_2^2 m_1^2 - 13 m_3^4 m_2^4 - 4 m_3^6 m_1^2 + 10 m_3^6 m_2^2 - 2 
         m_3^8) \Big] \ln{(m_2)}\nonumber\\
          &-  \Big[ p^8 - 2 (m_1^2 + 2 m_2^2 - 7 m_3^2) p^6 \nonumber \\&+ (13 
         m_1^4 - 34 m_2^2 m_1^2 - 13 m_2^4 - 22 m_3^2 m_1^2 + 16 m_3^2 m_2^2 + 26 m_3^4
         ) p^4 
         \nonumber \\&+ 2 (8 m_1^6 + 25 m_2^2 m_1^4 - 28 m_2^4 m_1^2 - 5 m_2^6 - 50 
         m_3^2 m_1^4 + 16 m_3^2 m_2^4 \nonumber \\&+ 11 m_3^4 m_1^2 - 8 m_3^4 m_2^2 + 7 m_3^6) p^2 
         \nonumber \\&- 
         ( - 32 m_2^2 m_1^6 + 26 m_2^4 m_1^4 + 4 m_2^6 m_1^2 + 2 m_2^8 + 16 m_3^2 
         m_1^6 + 50 m_3^2 m_2^2 m_1^4 - 56 m_3^2 m_2^4 m_1^2 
         \nonumber \\&- 10 m_3^2 m_2^6 - 13 m_3^4
          m_1^4 + 34 m_3^4 m_2^2 m_1^2 + 13 m_3^4 m_2^4 - 2 m_3^6 m_1^2 - 4 m_3^6 m_2^2
          - m_3^8) \Big] \ln{(m_3)} \Big\} \nonumber  \\
        &- \frac{1}{4\,p^2\,P(p^2,m_1,m_2,m_3)}  \Big\{
           (3 m_1^2 - 2 m_2^2 - 2 m_3^2) p^8 \nonumber \\&
         + 2 (2 m_2^2 m_1^2 - m_2^4 + 2 m_3^2 m_1^2 - 8 m_3^2 m_2^2 - m_3^4) p^6\nonumber \\&
         - 2 (5 m_1^6 - 8 m_2^2 
         m_1^4 + 8 m_2^4 m_1^2 - m_2^6 - 8 m_3^2 m_1^4 \nonumber \\&+ 2 m_3^2 m_2^2 m_1^2 + 5 m_3^2 
         m_2^4 + 8 m_3^4 m_1^2 + 5 m_3^4 m_2^2 - m_3^6) p^4 
         \nonumber \\&
         - 2 (4 m_1^8 - 6 m_2^2
          m_1^6 + 3 m_2^4 m_1^4 - m_2^8 - 6 m_3^2 m_1^6 + 8 m_3^2 m_2^4 m_1^2 
         \nonumber \\&+ 6 m_3^2 
         m_2^6 + 3 m_3^4 m_1^4 + 8 m_3^4 m_2^2 m_1^2 - 10 m_3^4 m_2^4 + 6 m_3^6 m_2^2 - 
         m_3^8) p^2 \nonumber \\&
         - (m_1 - m_2 - m_3) (m_1 - m_2 + m_3) (m_1 + m_2 - m_3) 
         (m_1 + m_2 + m_3) \nonumber \\& \times(m_1^2 - m_2^2 - m_3^2) (m_1^2 + m_2^2 + m_3^2) 
         m_1^2 \Big\} S^{(0)}(2;p^2)
          \nonumber \\
       &+ \frac{D(p^2,m_1,m_2,m_3)}{4\,p^2\,P(p^2,m_1,m_2,m_3)}   
           \left[ p^2 - (m_1^2 + m_2^2 + m_3^2) \right] S_1^{(0)}(2;p^2)  m_1^2 \nonumber\\   
       &-\frac{4}{p^2} \Big\{
           (p^2 + m_3^2) (m_1^2 - m_2^2) Z_2^{(1)}(2;p^2)
          + (p^2 + m_2^2) (m_1^2 - m_3^2) Z_3^{(1)}(2;p^2)  \Big\} \,, \labbel{eq:shiftS4}
\end{align}
where $P(p^2,m_1,m_2,m_3)$ and $D(p^2,m_1,m_2,m_3)$ are the usual polynomials
defined in Eq.s(\ref{defP},\ref{defD}).

In particular, from this equation we can read off the explicit values of the functions
$C(p^2,m_1,m_2,m_3)$, $C_1(p^2,m_1,m_2,m_3)$, 
$C_2(p^2,m_1,m_2,m_3)$ and $C_3(p^2,m_1,m_2,m_3)$ 
introduced in Eq.s(\ref{eq:tarinv},\ref{Im4SIm2a}):

\begin{align}
C(p^2,m_1,m_2,m_3) &= - \frac{1}{4\,p^2\,P(p^2,m_1,m_2,m_3)}  \Big[
           (3 m_1^2 - 2 m_2^2 - 2 m_3^2) p^8 \nonumber \\&
         + 2 (2 m_2^2 m_1^2 - m_2^4 + 2 m_3^2 m_1^2 
          - 8 m_3^2 m_2^2 - m_3^4) p^6\nonumber \\&
         - 2 (5 m_1^6 - 8 m_2^2 
         m_1^4 + 8 m_2^4 m_1^2 - m_2^6 - 8 m_3^2 m_1^4 \nonumber \\
         &+ 2 m_3^2 m_2^2 m_1^2 + 5 m_3^2 
         m_2^4 + 8 m_3^4 m_1^2 + 5 m_3^4 m_2^2 - m_3^6) p^4 
         \nonumber \\&
         - 2 (4 m_1^8 - 6 m_2^2
          m_1^6 + 3 m_2^4 m_1^4 - m_2^8 - 6 m_3^2 m_1^6 
          + 8 m_3^2 m_2^4 m_1^2 \nonumber \\&+ 6 m_3^2 
         m_2^6 + 3 m_3^4 m_1^4 + 8 m_3^4 m_2^2 m_1^2 
         - 10 m_3^4 m_2^4 + 6 m_3^6 m_2^2 - 
         m_3^8) p^2 \nonumber \\&
         - (m_1 - m_2 - m_3) (m_1 - m_2 + m_3) (m_1 + m_2 - m_3) 
         (m_1 + m_2 + m_3) \nonumber \\& 
         \times(m_1^2 - m_2^2 - m_3^2) (m_1^2 + m_2^2 + m_3^2) 
         m_1^2 \,\Big]\,, \labbel{defC}
\end{align}

\begin{align}
C_1(p^2,m_1,m_2,m_3) &=  
\frac{m_1^2\,D(p^2,m_1,m_2,m_3)}{4\,p^2\,P(p^2,m_1,m_2,m_3)}   
           \left[ p^2 - (m_1^2 + m_2^2 + m_3^2) \right] \,,  \labbel{defC1}
\end{align}

\begin{align}
C_2(p^2,m_1,m_2,m_3) = -\frac{4}{p^2} 
           (p^2 + m_3^2) (m_1^2 - m_2^2) \,, \labbel{defC2}
\end{align}

\begin{align}
C_3(p^2,m_1,m_2,m_3) = -\frac{4}{p^2} 
           (p^2 + m_2^2) (m_1^2 - m_3^2) \,. \labbel{defC3}
\end{align}

\section{Imaginary parts}
\labbel{App:im}
 
We work out here in some details the formulas used in Section \ref{imMI}. 
To start with, let us recall the definitions Eq.s(\ref{defbiR4}) 
\begin{align} 
 & (m_2-m_3)^2 = b_1 \le (m_2+m_3)^2 = b_2 \le (W-m_1)^2 = b_3 
                     \le (W+m_1)^2 = b_4 \ , \nonumber\\ 
 & R_4(b;b_1,b_2,b_3,b_4) = (b-b_1)(b-b_2)(b_3-b)(b_4-b) \ , 
\labbel{defbiR4a} \end{align} 
with 
\be W \ge (m_1+m_2+m_3) \ . \labbel{W>m_1+m_2+m_3} \ee 
Let us define 
\be I(n,W) = \int_{b_2}^{b_3} db\ b^n 
         \frac{1}{ \sqrtR } \ . \labbel{defIn} \ee 
One has obviously 
$$  \int_{b_2}^{b_3} db \frac{d}{db} 
         \Bigl[ b^n \sqrt{R_4(b;b_1,b_2,b_3,b_4)} \Bigr] = 0 \ ; $$ 
by working out the derivative, one gets an identity involving up to five 
integrals of the type $ I(n,W) $ with different values of $ n $; one finds 
that they can all be expressed as combination of four of them, which 
can be chosen to be 
\begin{align} 
 I(-1,W) &= \int_{b_2}^{b_3} \frac{db}{b\,\sqrtR}\,, \nonumber \\ 
 I(0,W) &= \int_{b_2}^{b_3} \frac{db}{\sqrtR}\,, \nonumber \\ 
 I(1,W) &= \int_{b_2}^{b_3} \frac{db\,\,b}{\sqrtR}\,, \nonumber \\ 
 I(2,W) &= \int_{b_2}^{b_3} \frac{db\,\,b^2}{\sqrtR}\,. \labbel{Inids} 
\end{align} 
In the same way, starting for instance from 
$$  \int_{b_2}^{b_3} db \frac{d}{db}
       \Bigl[ \frac{1}{b-b_1} \sqrt{R_4(b;b_1,b_2,b_3,b_4)} \Bigr] = 0 \ , $$
one finds 
\begin{align} \int_{b_2}^{b_3} \frac{db}{(b-b_1)\,\sqrtR} &= 
    \frac{1}{(b_2-b_1)(b_3-b_1)(b_4-b_1)} \nonumber\\ &\hspace{-3cm} 
 \times \bigl[ b_1(b_1-b_2-b_3-b_4)I(0,W) 
   + (b_1-b_2-b_3-b_4)I(1,W) - 2I(2,W) \bigr] \ . \labbel{R(b-b1)} 
\end{align}  
The above four integrals $ I(n,W), $ with $ n=-1,0,1,2 $, defined in 
Eq.s(\ref{Inids}), are easily 
expressed in terms of the usual complete elliptic integrals 
$ K(w^2), E(w^2), \Pi(a;w^2) $ of first, second and third kind, 
namely: \\ 
\begin{align} 
 K(w^2) &= \int_0^1 \frac{ dx}{ \sqrt{(1-x^2)(1-w^2\,x^2)} }\,, 
               \qquad 0<w^2<1\,, \labbel{defK} \\ 
 E(w^2) &= \int_0^1 dx \sqrt{\frac{1-w^2 x^2}{1-x^2}}\,, 
               \qquad 0<w^2<1\,,  \labbel{defE} \\ 
 \Pi(a;w^2) &= \int_0^1 
              \frac{dx}{ \sqrt{(1-x^2)(1-w^2\,x^2)}\ (1-a\,x^2) }\,,
              \qquad 0<w^2,a<1\,.  \labbel{defPi} 
\end{align} 
Indeed, the standard change of variable 
\be 
 b = \frac{ b_1 (b_3-b_2) x^2 - b_2 (b_3-b_1) } 
          { (b_3-b_2) x^2- (b_3-b_1) }\,, 
 \ \ \ x^2 =  \frac{(b_3-b_1)(b-b_2)}{(b_3-b_2)(b-b_1)} \ , \labbel{btox} 
\ee 
gives 
\begin{align}
 I(-1,W) &= \frac{2}{\sqrt{(b_3-b_1)(b_4-b_2)}}\frac{1}{b_1b_2}\, 
     \left[ b_2\ K(w^2) - (b_2-b_1)\ \Pi(a_1,w^2) \right]\,,  \nonumber\\ 
 I(0,W) &= \frac{2}{\sqrt{(b_3-b_1)(b_4-b_2)}}\, K(w^2)\,, \nonumber\\ 
 I(1,W) &= \frac{2}{\sqrt{(b_3-b_1)(b_4-b_2)}}\,\left[ \,b_1\ K(w^2) 
                       + (b_2-b_1)\ \Pi(a_1,w^2) \right]\,, \nonumber\\ 
 I(2,W) &= \frac{2}{\sqrt{(b_3-b_1)(b_4-b_2)}}\, 
           \left[ \,(b_1^2 + b_1 (b_2+b_3) - b_2 b_3)\ K(w^2) 
             - (b_3 - b_1)(b_4 - b_2)\ E(w^2) \right. \nonumber \\
        & \hspace{3.5cm} 
          \left. + (b_2-b_1)(b_1 + b_2 + b_3 + b_4)\ \Pi(a_2,w^2) \right]\,,
\labbel{ItoK} \end{align} 
where 
\begin{align} 
 w^2 &= \frac{(b_4-b_1)(b_3-b_2)}{(b_4-b_2)(b_3-b_1)} \ , \nonumber\\ 
 a_1 &= \frac{b_1(b_3-b_2)}{b_2 (b_3-b_1)}\ ,            \nonumber\\ 
 a_2 &= \frac{(b_3-b_2)}{(b_3-b_1)} \ . \nonumber 
\end{align} 
With the integral representation of Eq.s(\ref{defK},\ref{defE}) it is 
easy to obtain the formula 
\be \frac{d}{dw^2}K(w^2) = \frac{1}{2w^2} 
      \left[ \frac{E(w^2)}{1-w^2} - K(w^2) \right] \ , \labbel{derK} \ee 
which is useful for the evaluation of the imaginary parts of the 
Master Integrals $ S_i(d,p^2) $ in $ d=2 $ dimensions, Eq.(\ref{Im2Si}). \par 
One can easily express also $ K(w^2), E(w^2) $ etc. in terms of the 
$ I(n,W) $ by inverting 
Eq.s(\ref{ItoK}) or by using the change of variables Eq.(\ref{btox}). 
Indeed, the second of Eq.s(\ref{ItoK}) can also be written as 
\be K(w^2) = \frac{1}{2}\sqrt{(b_3-b_1)(b_4-b_2)}\ I(0,W) \ , 
             \labbel{KtoI0} \ee 
or, recalling the definition of $ I(0,W), $ Eq.(\ref{Inids}),  
\be \int_{b_2}^{b_3}  \frac{db}{\sqrtR} = 
      \frac{2}{\sqrt{(b_3-b_1)(b_4-b_2)}}\ K(w^2) \ , \labbel{KtoI0a} 
\ee 
while the change of variable $ x \to b $ gives 
$$ E(w^2) = \frac{1}{2}\sqrt{(b_3-b_1)(b_4-b_2)}\ \frac{b_2-b_1}{b_4-b_2} 
    \int_{b_2}^{b_3}  \frac{db}{\sqrtR}\ \frac{b_4-b}{b-b_1} \ , $$ 
which on account of Eq.(\ref{R(b-b1)}) can also be written as 
\be E(w^2) = - \frac{1}{2\sqrt{(b_3-b_1)(b_4-b_2)}} 
    \Bigl[ (b_2 b_3 + b_1 b_4)\ I(0,E) 
    - (b_1 + b_2 + b_3 + b_4)\ I(1,E) + 2\ I(2,E) \Bigr]\ . \labbel{EtoIi} 
\ee 
\par 
Let us discuss shortly also the limit of equal masses $ m_1 =m_2 =m_3 =m, $ 
which gives $ b_1 = 0, b_2=4m^2, b_3=(W-m)^2, b_4=(W+m)^2. $ In that 
limit, thanks to $ b_1=0 $ we can read Eq.(\ref{R(b-b1)}) as an identity 
expressing $ I(-1,W) $ in terms of $ I(1,W), I(2,W) $. Further, 
one more identity appears \cite{RBER1975} from 
$$ \int_{4 m^2}^{(W-m)^2} db \, \frac{d}{db} \, 
    \ln{ \left( \frac{ b (W^2+3 m^2-b) 
                         + \sqrt{R_2(b,m^2,m^2)}\sqrt{R_2(W^2,b,m^2)} } 
                     { b (W^2+3 m^2-b) 
                         - \sqrt{R_2(b,m^2,m^2)}\sqrt{R_2(W^2,b,m^2)} } 
          \right) } = 0 \ , $$ 
where $ R_2(a,b,c) $ is defined in Eq.(\ref{defR2}), giving (in 
the equal mass limit), 
\be I(1,W) = \frac{1}{3}(W^2 + 3 m^2) \ I(0,W) \ , \labbel{I1toI0} \ee 
showing once more that in the equal mass limit the imaginary parts 
can be expressed in terms of two independent functions only. 

 
\end{document}